\documentclass[conference, 10pt]{IEEEtran}
\usepackage{citesort}
\usepackage{color}
\usepackage{framed}
\usepackage{url}
\usepackage{verbatim}
\usepackage{caption}
\usepackage{graphicx}
\usepackage{amsmath, amssymb, amsfonts, amsthm} 
\usepackage{epsfig}
\usepackage{mathtools}
\usepackage{url}
\usepackage{algorithm}
\usepackage[noend]{algpseudocode}
\usepackage{subfigure}
\usepackage{wrapfig}
\usepackage{multirow}
\usepackage{enumitem}
\usepackage{tikz}
\usepackage{pgfplots}
\newenvironment{myquote}%
  {\smallskip \list{}{\leftmargin=0.2in\rightmargin=0.2in}\item[] \em}%
  {\endlist}

\newtheorem{theorem}{Theorem}

\renewcommand{\paragraph}[1]{\smallskip \noindent {\bf #1.}~} 

\newcommand{\BT}{\begin{theorem}}
\newcommand{\ET}{\end{theorem}}

\newcommand{\BD}{\begin{definition}}
\newcommand{\ED}{\end{definition}}

\newcommand{\BCR}{\begin{corollary}}
\newcommand{\ECR}{\end{corollary}}

\newcommand{\BEX}{\begin{example}}
\newcommand{\EEX}{\end{example}}

\newcommand{\BL}{\begin{lemma}}
\newcommand{\EL}{\end{lemma}}

\newcommand{\BP}{\begin{proposition}}
\newcommand{\EP}{\end{proposition}}

\newcommand{\BCM}{\begin{claim}}
\newcommand{\ECM}{\end{claim}}

\newcommand{\BPF}{\begin{proof}}
\newcommand{\EPF}{\end{proof}}
\newcommand{\BEN}{\begin{enumerate}}
\newcommand{\EEN}{\end{enumerate}}

\newcommand{\BI}{\begin{itemize}}
\newcommand{\EI}{\end{itemize}}

\newcommand{\BO}{\begin{observation}}
\newcommand{\EO}{\end{observation}}

\newcommand{\BDS}{\begin{description}}
\newcommand{\EDS}{\end{description}}

\def\squareforqed{\hbox{\(\blacksquare\)}}
\def\qed{\hfill \squareforqed}

\newcommand{\etal}{{\it et al }}

\newcommand{\secparam}{\ensuremath{\lambda}}

\newcommand{\ignore}[1]{}

\newcounter{defcounter}
\newlength{\protowidth}

\newcommand{\olrk}[1]{%
   \ifx\nursymbol#1\else\!\!\mskip4.5mu plus 0.5mu\left(#1\right)\fi}
\newcommand{\elrk}[1]{%
   \ifx\nursymbol#1\else%
        \!\!\mskip4.5mu plus0.5mu\left[\mskip2.5mu plus0.5mu #1\right]\fi}

\def\path{\ensuremath{{\sf location}}}

\def\Access{\mathsf{PhysicalAcc}}

\def\Access{\ensuremath{{\sf Access}}}

\def\O{\mathbb{O}}

\graphicspath{{./figures/}}

\newcommand{\sysname}[1]{ConcurORAM}{\ignorespaces}
\newcommand{\order}[1]{O(#1)}

\newtheorem{thm}{Theorem}
\newenvironment{customthm}[1]
  {\renewcommand\thethm{#1}\thm}
  {\endthm}

\newtheorem{definition}{Definition}

\def\etal{et\ al.~}

\title{\huge ConcurORAM: High-Throughput Stateless Parallel Multi-Client ORAM}
\begin{document}

\author{\IEEEauthorblockN{Anrin Chakraborti\IEEEauthorrefmark{1},
		Radu Sion\IEEEauthorrefmark{1}
	}
	
	\IEEEauthorblockA{\IEEEauthorrefmark{1}Stony Brook University, 
		\{anchakrabort, sion\}@cs.stonybrook.edu\\}

}

\maketitle
\begin{abstract}
	\sysname~ is a parallel, multi-client oblivious RAM (ORAM) that eliminates
	waiting for concurrent stateless clients and allows overall throughput to scale gracefully, without requiring trusted third party components (proxies) or direct inter-client coordination.
	A key insight behind \sysname~ is the fact that, during multi-client data
	access, only a subset of the concurrently-accessed server-hosted data
	structures require access privacy guarantees.  Everything else can be safely
	implemented as oblivious data structures that are later synced securely and
	efficiently during an ORAM ``eviction''. 	
	Further, since a major contributor to latency is the eviction -- in
	which client-resident data is reshuffled and reinserted back encrypted into
	the main server database -- \sysname~ also enables multiple concurrent
	clients to evict asynchronously, in parallel (without compromising
	consistency), and in the background without having to block
	ongoing queries. 
	

	
	%
	
	
	As a result, throughput scales well with increasing number of concurrent
	clients and is not significantly impacted by evictions.  For example, about
	65 queries per second can be executed in parallel by 30 concurrent clients,
	a 2x speedup over the state-of-the-art \cite{taostore}.  The query access time for individual clients 
	increases by only 2x when compared to a single-client deployment.
	
	
\end{abstract}
\section{Introduction}
\label{oram:intro}

%
As increasing amounts of confidential data are outsourced in today's
cloud-centric environments, providing confidentiality and privacy becomes
critical.  To ensure confidentiality, outsourced data and associated
metadata can be encrypted client-side.  Data remains encrypted throughout
its lifetime on the server and is decrypted by the client upon retrieval. 
However, encryption is simply not enough for ensuring confidentiality
since {\em access patterns} may leak significant
information \cite{accesspatternleak}.

Oblivious RAM (ORAM) allows a client to hide data access patterns from an
untrusted server hosting the data.  Informally, the ORAM adversarial model
ensures indistinguishability between multiple equal-length client query
sequences.  Since the original ORAM construction by Goldreich and Ostrovsky
\cite{goldreich}, a large volume of literature
\cite{bforam,privatefs,pathoram,ringoram}
has been dedicated to developing more efficient ORAM constructions.
Of these, under an assumption of $\O(n)$ client storage with small
constants, PathORAM \cite{pathoram} is widely accepted as asymptotically the
most {\em bandwidth efficient} ORAM.  RingORAM \cite{ringoram} further
optimizes PathORAM for practical deployment by reducing the
bandwidth complexity constants.
%

%
Although tree-based ORAM designs \cite{pathoram,ringoram} have achieved near-optimal {\em bandwidth}
for single-client scenarios, one critical challenge remains un-addressed,
namely the ability to accommodate multiple concurrent clients efficiently.
It is straight-forward to deploy existing schemes to support multiple
clients by sharing ORAM credentials and storing data structures that
would normally be maintained client-side (e.g., the stash and the position
map in the case of a tree-based ORAM) on the server to ensure state
consistency across multiple clients.  However, in such a setup, to maintain
access privacy, only {\em one client} can be allowed to access the
server-hosted data structures at any one time. This reduces the overall throughput and significantly increases the query response time.  A client may need to wait
for {\em all} other clients to finish before retrieving a data item.  Since
ORAMs often have non-trivial query latencies, this usually results in significant access latency for a client before being able to proceed with the query.

An existing line of work on parallel ORAM constructions
\cite{chenpram,elletepram,katzopram,chanopram,circuitOPRAM} achieve parallelism at no additional bandwidth
cost under the assumption of constant inter-client awareness and
communication. Although a
step forward, this poses barriers that are often times difficult to handle
in real scenarios.
Without inter-client communication, Taostore~\cite{taostore} assumes a
trusted close-to-server (proxy) to achieve parallelism.  All client requests
are routed to the proxy, which deploys multiple threads to fetch one (or
more) paths from a PathORAM~\cite{pathoram} data tree and satisfy numerous
client requests at once.  The need for a trusted third-party however
deviates from the standard ORAM model, where trust is only placed on the
clients at most.  Moreover, a trusted proxy may be difficult to deploy in
reality as well as presenting a single point of failure/compromise. 
%

\paragraph{Motivation}
This paper addresses these shortcomings by eliminating the need for trusted
proxies and inter-client communication, and allowing client queries to
proceed {\em independently} in the presence of other ongoing queries.  {\em
Thus, \sysname~ is the first to achieve parallelism for stateless ORAM clients in the standard trust model without the need for direct inter-client
communication}.

\subsection{Challenges and Key Insights}


\paragraph{Asynchronous Accesses}
Tree-based ORAMs feature two different classes of accesses to the server:
(i) {\em queries} (reading a root-to-leaf path) and {\em evictions} (writing
back some of the previously read data items to the root-to-leaf path).  A
common strategy for reducing overall bandwidth/round-trips is to couple
queries and evictions \cite{pathoram} .  Even if a multi-client
design can be envisioned in which metadata is stored on the server for
consistency, this coupling forces a synchronous design in which only one
client is in charge at any given time.

One approach for decoupling accesses is to maintain and update a
locally-cached subtree and smartly sync with on-server data
structures without blocking queries \cite{taostore}.  Since, the local
subtree essentially performs the role of a write-back cache, the
construction does not immediately scale to a multi-client setting.
This can be resolved by having a designated trusted client (a
close-to-server proxy) maintain the subtree locally and route queries to the
server. However, the assumption of a trusted proxy introduces several performance and security drawbacks. The most important of these is that the system's overall performance now entirely depends on the proxy's resources e.g., available network bandwidth. An under-provisioned proxy or a system failure/compromise will adversely affect all clients in the system. 
Also, this design does not support stateless clients (or storage-limited clients) -- the proxy carefully synchronizes accesses based on (potentially large amount of) metadata stored locally. Outsourcing this metadata naively may lead to privacy leaks.

To eliminate this security/performance bottleneck and allow practical
interactions with asynchronous decoupled multi-client operations, \sysname~
adopts a different approach.  Queries only perform non-blocking read-only
accesses.  Further,
evictions write back changes with access privacy in the background to
additional server-hosted oblivious data structures (ODS) (append-only logs,
write-only tree etc.).  These data structures are designed to be
synchronized with the main ORAM tree periodically and efficiently.  The
synchronization mainly involves copying contents between server-hosted data
structures, reference swaps etc. which can be performed securely and efficiently with limited client interaction, optimizing both bandwidth and round-trips.


\paragraph{Parallel Queries}
Asynchronous accesses alone cannot facilitate parallel query execution.  The
query protocols for tree-based ORAMs are fairly complex and involve multiple
read/write-back steps.  Without careful synchronization, overlapping
accesses will violate consistency and privacy.

Possible solutions are parallel query abstractions such as in PrivateFS
\cite{privatefs}, which protects inter-client query privacy for any
multi-client ``non-simultaneous'' ORAM.  As we will see, in \sysname~, this
forms the basis of a parallel sub-query mechanism, suitably modified to
support parallel evictions.

\paragraph{Parallel Evictions} 
While parallel query execution is a good starting point, evictions are
at least as expensive as the queries.  Thus, even with
parallel queries, serialized evictions will be the de facto bottleneck and limit 
overall throughput gains.

\sysname~ overcomes this limitation by allowing multiple evictions to
execute in parallel.  In essence, this is possible because we observe that
when evictions are performed according to a {\em deterministic eviction
schedule} \cite{ringoram}, and the number of parallel evictions is fixed,
access patterns to the server-hosted data structures can be clearly and
deterministically defined.

This allows identifying the critical sections of the eviction protocol where
synchronization is necessary, and the design of associated fine-grained
locks.  The remainder of the protocol can be performed in parallel using additional server-hosted data structures designed to maintain state and enforce
minimally-sufficient global synchronization.

\paragraph{Evaluation} 
As we will see, because the critical sections are small, this results in an
overall throughput that scales gracefully with increasing number of
concurrent clients and is not significantly impacted by evictions.  For
example, about 65 queries per second can be executed in parallel by 30
concurrent clients with only a 2x increase in query access time over a
single-client deployment. Importantly, this is a 2x speedup over the state-of-the art \cite{taostore}, which operates under stronger assumptions of a trusted proxy.

\section{Related Work}
\label{oram:related}

ORAMs have been well-researched since the seminal work by Goldreich and Ostrovsky \cite{goldreich}. 
We specifically discuss existing parallel ORAM constructions here and refer to the vast amount of existing literature for further details on general ORAM construction \cite{goldreich,privatefs,bforam,pathoram,ringoram}. 

\paragraph{Oblivious Parallel RAM (OPRAM)}
Boyle {\em et al.} \cite{elletepram} first introduced an oblivious parallel RAM (OPRAM) construction assuming inter-client communication for synchronization. Clients 
coordinate with each other through an {\em oblivious aggregation} operation and 
prevent simultaneous clients from querying for the same block. For colliding client accesses, 
only {\em one representative} client queries for the required item while 
all other clients query for dummy items. The {\em representative} client then communicates 
the read item to all other colliding clients through an {\em oblivious multi-cast} operation.
Subsequent works \cite{chanopram,katzopram,chenpram,opram_chan_asiacrypt,circuitOPRAM} have optimized Parallel RAMs matching the overhead of a sequential ORAM construction.

\paragraph{TaoStore \cite{taostore}}
Another interesting parallel ORAM construction is TaoStore 
which achieves parallelism for PathORAM. 
through a trusted proxy. All client 
queries are redirected to the trusted proxy which then queries for the 
corresponding paths from the PathORAM data tree. Further, the 
proxy runs a secure scheduler to ensure that the multiple path reads do not overlap and leak 
correlations in the underlying queries. TaoStore 
achieves a significant increase in throughput but can support only a 
limited number of parallel clients before the throughput plateaus 
due to the proxy's bandwidth constraints.

\paragraph{PD-ORAM \cite{privatefs}}
Williams \etal provided a parallel ORAM construction that does not require trusted proxies and inter-client communication. However, the construction  is derived from a hierarchical ORAM construction, with higher access complexity than standard tree-based ORAMs. Hence, the overall throughput gain is limited. 

\section{Background}
\label{model}

\subsection{Oblivious RAM}
 Oblivious RAM (ORAM) is a cryptographic primitive that allows a client/CPU to 
 hide its data access patterns from an untrusted server/RAM hosting the accessed 
 data. Informally, the ORAM adversarial model prevents an adversary from 
 distinguishing between equal length sequences of queries made by the client to 
 the server. This usually also includes indistinguishability between reads and 
 writes. We refer to prior works for more formal definitions \cite{bforam,privatefs,pathoram,ringoram}.

\subsection{PathORAM}
\label{backround:pathoram}
PathORAM is an efficient ORAM construction with an overall query asymptotic access complexity of $\O(\log N)$ blocks, matching the known lower bound \cite{goldreich}.
PathORAM organizes data as a binary tree. Each node of the tree is a {\em bucket} with multiple (constant number of) blocks. 
A block is randomly mapped to a unique path in the tree.

\begin{myquote}
Invariant: A block mapped to a path
resides either in any one of the buckets on the 
path from the root to the corresponding leaf, or in a {\em stash} that is stored locally
\end{myquote}

\paragraph{Position Map}
PathORAM use a ``position map'' data structure to map
logical data item addresses to identifiers of tree leafs defining a corresponding path
from the root, ``within'' which the data items are placed. Specifically, a
data item ``mapped'' to leaf ID $l$ can reside in any of the nodes along the path from the root to leaf $l$. The position map is either stored on the client ($\O(N)$ client storage) or recursively in smaller ORAMs on the server.

\paragraph{Access}
To access a particular block, the client downloads all the contents along the root-to-leaf path
to which the block is mapped. Once the block has been read, it is remapped to a new leaf and {\em evicted} 
back to the tree. Various eviction procedures have been proposed in literature \cite{pathoram,ringoram}. 
We specifically describe the RingORAM~\cite{ringoram} protocol as a building block.

\subsection{RingORAM}
\label{background:ringoram}
RingORAM~\cite{ringoram} is derivative of PathORAM \cite{pathoram} that optimizes practical bandwidth 
requirements. This is the result of two optimizations: i) de-coupling the queries from the expensive eviction procedure, 
and ii) fetching only {\em one} block from each bucket in the tree during queries.

Unlike PathORAM \cite{pathoram}, where a query 
needs to fetch all buckets along a path from the root to a particular leaf, RingORAM query cost is independent of the bucket size. This is achieved by storing additional dummy blocks in each bucket. Bucket-specific metadata tracks the locations of 
blocks (and dummy blocks) within buckets. Each query first reads this 
metadata and determines whether the required block is present in a particular bucket. A dummy block is fetched from the buckets 
that do not contain the required data block.

The additional dummy blocks makes the RingORAM buckets larger than PathORAM buckets. This makes evictions expensive. To overcome this, 
RingORAM delays evictions by de-coupling queries and evictions -- an eviction is performed after a fixed number of queries. To make evictions more effective, Ring ORAM uses a deterministic eviction schedule based on the reverse-lexicographical ordering of leaf IDs to select eviction paths.

\paragraph{Query} The query protocol in RingORAM is as follows

\begin{enumerate}
	\item Determine the path to which the block is mapped using the position map.
	\item For each bucket on the path
	\begin{itemize}
		\item Use the bucket-specific metadata to determine if the required block is in that bucket.
		\item If the block is in the bucket, read the block. Otherwise, read a dummy block.
	\end{itemize}
	\item Download the entire stash (if stored on the server). Add the queried block to the stash (if not already present).
	\item Remap the block to a new randomly selected path, update position map accordingly. 
\end{enumerate}

\paragraph{Evictions}After a fixed number of queries, an eviction is performed as follows

\begin{enumerate}
	
	\item Download the contents of an entire path from the tree determined by the {\em reverse-lexicographical ordering of the leaf IDs}. Place the contents in the stash.
	\item Write back as many blocks as possible from the stash (re-encrypted) to a new locally created path.
	\item Write back the contents of the new path to the tree
	
\end{enumerate}

\paragraph{Deterministic Selection of Eviction Paths}
RingORAM selects eviction paths by ordering the leaf identifiers in the {\em reverse-lexicographical order}. Specifically, the reverse lexicographical representation denotes each path of the data tree as a {\em unique} binary
string.  The least significant bit (LSB) of the string assigned for a path
is 0 if the target leaf corresponding to the path is in the left subtree of
the root and 1 otherwise.  The process is continued recursively for the next bits, up to the leaf, with the next bit(s) being assigned based on whether the leaf is in the left or right subtree of the children nodes. 

Intuitively this results in better evictions by spreading out blocks uniformly across the tree since consecutive eviction paths have minimum overlaps, and $N$ paths of the tree are selected once before the same path is selected again.  

\paragraph{Access Complexity}
ORAMs are typically evaluated in terms of {\em bandwidth} -- the number of {\em data blocks} that are downloaded/uploaded in order to 
complete one logical request. RingORAM features an overall bandwidth of $\O(\log N)$ data blocks, where $N$ is the total number of blocks in the ORAM. This asymptotic bound holds only under the {\em large block size assumption} when the data blocks size is $\Omega(\log^{2} N)$ bits. 

{\sysname~} has the same large block size assumption and all access complexities reported in this paper  indicate the number of physical blocks that are accessed overall for a fulfilling a particular logical request.

%

\section{Overview}
\label{overview}

\subsection{Security Definitions}

\paragraph{Trust Model}
There are two types of parties: the ORAM clients (with limited local
storage) and the ORAM server (a remote storage hosting client data).

\begin{itemize}
	
	\item {\em Honest-but-curious server}: The server can observe all
	requests and attempts to correlate them by saving and comparing snapshots. 
	The server does not deviate from the \sysname~ protocol.
	
	\item {\em Trusted clients}: Clients are {\em honest} and share
	secrets (credentials, keys, hashes etc.) required for accessing the ORAM. 
	{\em Clients do not need to interact with each other, but can observe and track other client accesses through the server-hosted data structures.} 
	
	
\end{itemize}

The security model for parallel ORAMs followed in this work closely resembles the model proposed in \cite{privatefs}. 
First, note that security for non-parallel ORAMs is defined in terms of query privacy.

\begin{definition}[Query Privacy] 
	\label{def_single}
	Let $\vec{y} = (y_1, y_2, \ldots)$ denote a sequence of
	queries, where each $y_i$ is a ${\sf Read}(a_i)$ or a
	${\sf Write}(a_i, d_i)$; here, $a_i \in [0, N)$ denotes the logical address
	of the block being read or written, and $d_i$ denotes a block of data being
	written. For an ORAM scheme $\Pi$, let $\Access^\Pi(\vec{y})$ denote the
	physical access pattern that its query protocol produces for the logical
	access sequence $\vec{y}$.  We say the scheme $\Pi$ is {\em secure} if for any
	two sequences of queries $\vec{x}$ and $\vec{y}$ of the same length, it
	holds
	$$ \Access^\Pi(\vec{x}) ~~ \approx_c ~~ \Access^\Pi(\vec{y}), $$
	where $\approx_c$ denotes computational indistinguishability (with respect to
	the security parameter $\secparam$).
	
\end{definition}

In other words, access transcripts produced by queries are indistinguishable 
and independent of the items being accessed. 

\paragraph{Parallel ORAM Requirements}
Boyle \etal \cite{elletepram} formally define the properties required for a parallel ORAM compiler. Our goals are similar and we informally restate these requirements in the client-server setting

\begin{itemize}
	\item {\em Correctness:} Given a set of parallel queries $(q_1, q_2, \ldots)$ for items $(i_1, i_2, \ldots)$, the query protocol should return the most up-to-date versions of $(i_1, i_2, \ldots)$ with very high probability.
	
	\item {\em Obliviousness:} 	Let $\vec{y} = (y_1, y_2, \ldots)$ denote a set of queries from multiple parallel clients, where each $y_i$ is a ${\sf Read}(a_i)$ or a
	${\sf Write}(a_i, d_i)$; here, $a_i \in [0, N)$ denotes the logical address of the block being read or written, and $d_i$ denotes a block of data being
	written. For an ORAM scheme $\Pi$, let $\Access^\Pi(\vec{y})$ denote the
	physical access pattern that its query protocol produces when the queries 
	in $\vec{y}$ are executed in parallel. We say the scheme $\Pi$ is {\em secure} if for any 	two sequences of parallel queries $\vec{x}$ and $\vec{y}$ of the same length, it
	holds
	$$ \Access^\Pi(\vec{x}) ~~ \approx_c ~~ \Access^\Pi(\vec{y}), $$
	where $\approx_c$ denotes computational indistinguishability (with respect to
	the security parameter $\secparam$).

\end{itemize}

\paragraph{Query-Privacy Inheritance}
In the case of a parallel ORAM (without inter-client communication), the server observes queries from multiple parallel clients. To ensure obliviousness, as a first step, all clients must follow the same query protocol. Further, a particular requirement in the multi-client case is that multiple clients cannot query for the same item at the same time -- only one client queries for the real item while all other clients must issue ``fake queries''. The fake queries should be indistinguishable from real queries. This corresponds to a case in the serial ORAM where a particular item is found in the local client-side cache and the client generates a fake query instead to maintain obliviousness. As we will see, this is ensured by \sysname~.

When these requirements are met, the combined physical access trace generated by executing queries in parallel will simply be an interleaving of the access traces generated by the individual queries when executed serially \cite{privatefs}. For parallel ORAMs that follow this model, Williams and Sion \cite{privatefs}  define and prove the following property

\begin{definition}[Query Privacy Inheritance]
	\label{def:query_privacy_inheritance}
	If there exists an adversary with non-negligible advantage at violating query privacy in a parallel ORAM, then by implication there exists an adversary with non-negligible advantage at violating query privacy in the underlying non-parallel single-client ORAM.
\end{definition}

The intuition here is that given a set of parallel queries (which may include overlapping items), if the queries can be executed serially without violating query privacy, then the queries can also be executed in parallel by interleaving the accesses without violating query privacy. The combined access traces of the interleaved queries in the parallel case, does not provide any extra information to the adversary than what is already available due to the individual query access patterns. Thus, to show obliviousness for in the parallel case, we must show that the underlying ORAM query protocol ensures query-privacy in the serial case (Definition \ref{def_single}).

\subsection{Preliminaries}
\label{overview:preliminaries}

\paragraph{Server Storage} As with most tree-based ORAMs, the main server-side
data structure is a binary tree storing fixed-sized data blocks. 
Specifically, a database with $N$ logical blocks requires a binary tree
with $N$ leaves.

\paragraph{Node Structure} {\sysname~} follows the same node structure as Ring ORAM \cite{ringoram}. Specifically, each node of the tree contains a fixed number of data blocks (denoted by $Z$) and dummy blocks (denoted by $S$), collectively referred to as a bucket. Note that $(Z+S) \in \O(\log N)$ blocks.
Blocks in a bucket are encrypted
with semantic security (e.g., using randomized encryption) and randomly
shuffled.  Buckets also store relevant metadata for retrieving specific data
blocks, identifying dummy blocks etc.

\paragraph{Stateless Clients}
In addition to the main data tree, tree-based ORAMs require several
auxiliary data structures for maintaining state.  This includes a {\em
	position map} to track the locations of logical blocks in the ORAM data tree
and a {\em stash} to hold blocks that could not be immediately placed back
on the tree (due to randomized placement).  Typically, these are stored
client-side to speed-up accesses.  However, in a multi-client setting, this
metadata needs to be stored persistently on the server, in order to present
a consistent view for all clients.

\paragraph{Temporary Client Storage} Clients require a small amount of temporary storage to perform
several operations locally before uploading updates to the server.  This
storage is bounded by $\O(\log N)$ block for a database with $N$ blocks.

\paragraph{Building Blocks}
For illustration purposes, we will consider the RingORAM query and eviction
protocols (described in Section \ref{background:ringoram}) as a
(non-parallel) starting point for some of the \sysname~ protocols.

We note however that this is not necessary -- the techniques presented here
can be generalized for other tree-based ORAMs provided the following two
conditions are satisfied 

\begin{itemize}
	\item The ORAM supports {\em dummy queries} (where only dummy blocks are fetched from the tree), and the dummy queries are indistinguishable from real queries.
	\item Eviction paths are selected using the reverse-lexicographical ordering of leaf identifiers.	
\end{itemize}

\paragraph{Position Map Design}
For a concurrent ORAM design, the position map itself needs to allow concurrent access 
to position map entries. In {\sysname~}, this can be realized by storing the position map recursively in smaller {\sysname~} instances. However, this introduces several implementation challenges due to concurrent recursive data structure accesses.

To avoid recursion, a simpler alternative is to store the entire position map in a parallel hierarchical ORAM e.g. PD-ORAM \cite{privatefs}, which does not require its own position map. In the current design and implementation of  {\sysname~},  the position
map is stored in a PD-ORAM instance shared by the clients. 
Since PD-ORAM supports parallel multi-client accesses, the position map is treated as a secure black box.
%


%

We specifically note that this approach does not affect overall complexity. Position map entries for $N$ blocks are typically $\O(\log N)$ bits in size. The overall PD-ORAM access complexity is $\O(\log^{2}N)$ items. Thus, accessing a position map entry from PD-ORAM has an access complexity of $\O(\log ^{3} N)$ bits, or $\O(\log N)$ blocks of size $\Omega (\log^{2}N)$ bits.  This is equivalent to a data access for tree-based ORAMs.

\paragraph{Server Functionalities}
Without the aid of explicit inter-client communication or trusted proxies,
achieving a notion of global synchronization in \sysname~ requires
server-managed mechanisms and APIs for fine-grained locking, all while
guaranteeing access privacy and obliviousness. 

This is different from prior work \cite{taostore} where inter-client
synchronization is handled by the proxy and the server is treated basically
as a storage device, only providing APIs for downloading,
uploading, copying and deleting data.

\begin{figure}[t]
	\centering
	\includegraphics[scale=0.15]{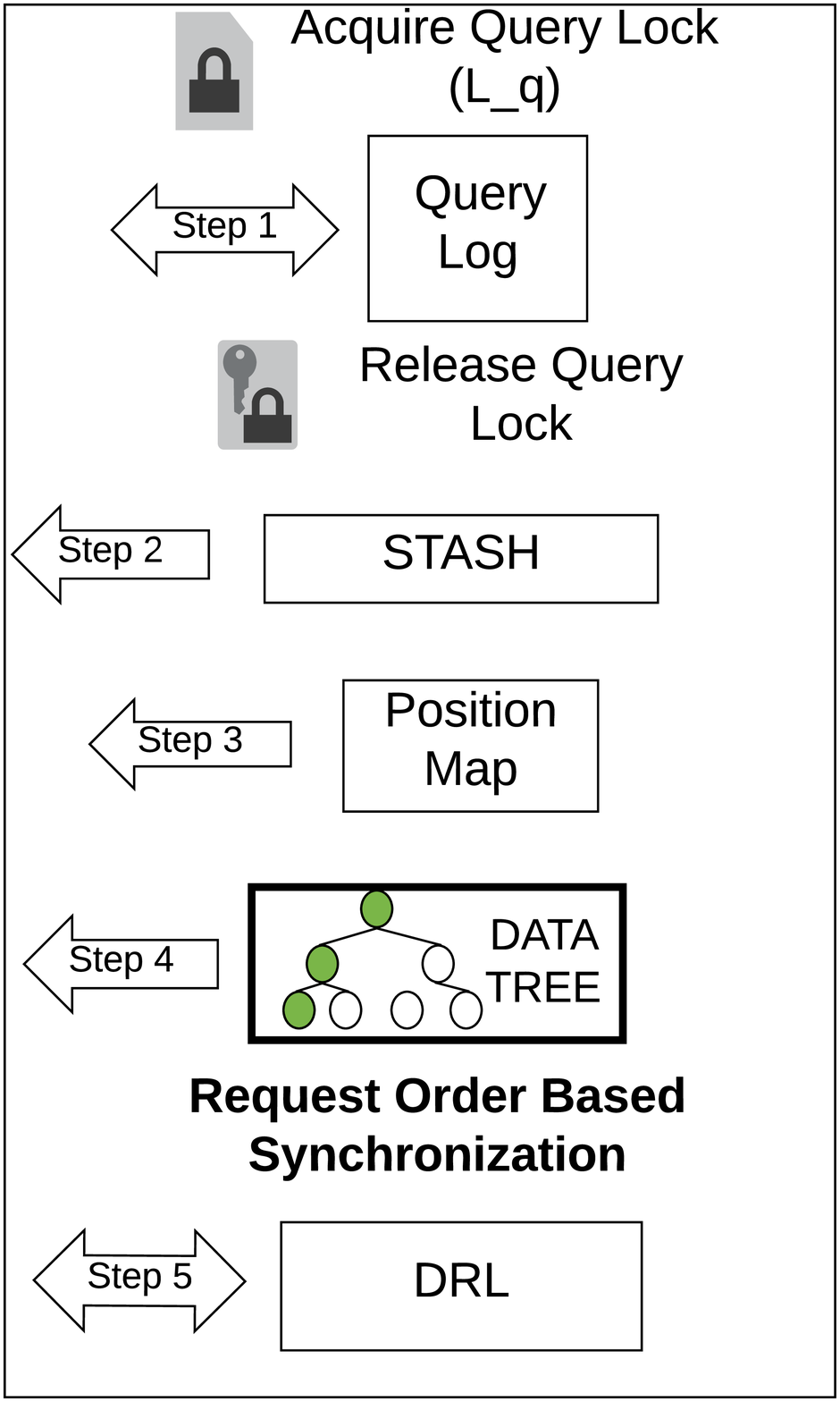}
	\vspace{-0.5cm}
	\caption{
		\footnotesize
		Parallel query overview.  Accesses to the stash, the
		position map and the data tree can proceed
		concurrently.  Queries access the query log after
		acquiring a mutex (the query lock) for a short
		period of time.  Synchronizing accesses to the DRL
		ensures that queries are completed in the order in
		which they start execution, a necessary condition to
		prevent security leaks.\label{query_overview}
	}
	\vspace{-5pt}
\end{figure}

\subsection{Parallel Queries}
\label{overview:parallel_queries}
Supporting parallel queries is the first step towards achieving full
parallelism.  On observation, it may be evident that if two clients
concurrently query for the same block, according to the query protocol in
Section \ref{background:ringoram}, they will access the same path from the
tree, thereby leaking inter-client query privacy.  Prior work solves this by
either assuming direct inter-client communication and synchronization
\cite{elletepram,chenpram} or by routing queries through a common proxy
\cite{taostore} which executes only the non-overlapping queries.

Without requiring these assumptions, \sysname~ provides inter-client
awareness through server-hosted data structures.  The techniques presented
here are similar to the parallel query abstraction described in
\cite{privatefs} -- the goal is to convert an ORAM that is secure for
non-parallel queries into an ORAM with parallel queries (Figure \ref{query_overview}), augmented with
judiciously designed server-hosted data structures.




\paragraph{Query Log} 
To support concurrent queries without leaking inter-client privacy,
information about all ongoing transactions is written to an encrypted {\em query log}, not unlike a transaction log. 
Prior to executing a query, clients first download the entire log (to check
for overlapping accesses) and then append to the log the encrypted
logical address of the data block they are querying for.
This ensures that all clients have a consistent view of ongoing
transactions.  In case of an overlapping query -- when there is a previous
entry for the same block in the query log -- the client proceeds with a {\em
	dummy query} by simply reading dummy blocks from a random path in the tree.

\paragraph{Data Result Log (DRL)} 
For overlapping queries, the above ensures that only {\em one} client can
access the target block at any time.  The other clients must wait for an
eviction to re-randomize the location of the block in the tree before they
will access the block.  However, this may result in indefinite wait times
and possibly leak privacy under a timing channel \cite{taostore}.

To mitigate, \sysname~ caches previously accessed blocks in a {\em data
	result log} until {\em periodic} ORAM evictions can place them back to
random locations on the data tree.  

At the end of an access, the block
queried by a client is re-encrypted and appended to the DRL.  Other clients
that queried for the same block (but ended up performing a dummy query
instead) can then access the block by reading the entire DRL.


\paragraph{Request Order-based Synchronization} 
Since clients do not communicate, it is not possible for a particular client
to learn when its target block is in the DRL.  Further, if clients accessed
the DRL {\em only} in case of dummy queries, the server will be able to
distinguish overlapping queries for the same block.
The following DRL access protocol resolves all these concerns

\begin{enumerate}
	
	\item Clients executing a query read the DRL after all ``previous''
	clients finish their queries.  Specifically, queries that started execution
	earlier by registering an entry in the query log, must complete {\em all}
	steps of the query protocol before the current client can read the DRL.  The
	client checks this by comparing the size of the current DRL and the ordering
	of entries in the query log.  If the client registered the $i^{th}$ entry to
	the query log, it can proceed only after there are $i-1$ blocks in the DRL. Note that if a client executed a dummy query, then it will necessarily find the target block in the DRL after all previous queries have finished execution.

	\item After reading the entire DRL, a client always appends the
	(updated and re-encrytped) target block to the DRL.
	
\end{enumerate}

\paragraph{Query Round \& Log Size}
To bound the log sizes, 
{\sysname~} introduces the notion of a {\em query round}. Specifically, only up to $c$ (a constant) queries 
are allowed to execute in parallel, while requests that arrive later have to wait until all executing queries finish -- this includes appending their results to the data result log. A round of $c$ parallel queries constitutes a query round in \sysname~. Queries arrive and execute individually after registering an entry in the query log, but belong to the same query round as long as the current query log contains less than $c$ entries. 
Once all queries in the current query round finish execution, the contents of the DRL (which now contains $c$ blocks) are evicted back to the data tree and the DRL and the query log are cleared. This ensures that the query log and DRL size never exceed $c$ entries.


\subsection{Non-blocking Evictions}
\label{overview:non_blocking}
Parallelizing the query step alone is not sufficient to achieve good
scalability.  Evictions are often expensive and can block clients for
possibly impractical amounts of time.
Instead, to scale, \sysname~ performs evictions continuously in the
background ensuring that queries are blocked for very short
upper-bound periods of time.  Achieving this is not straightforward. We need to introduce several
key insights. 

First, note that in the non-parallel case,
queries and evictions (Section
\ref{background:ringoram}) include the following client-side steps

\begin{itemize}
	\item {\em Eviction}
	\begin{enumerate}
		\item Fetch buckets from a path of the data tree.			
		\item Evict contents of the stash to new path locally. 
		\item Write back the new path and stash.
		\item Update the position map and clear the logs.
	\end{enumerate}
	
	\item {\em Query}
	\begin{enumerate}
		\item Update the query log.
		\item Read the stash and position map query.
		\item Fetch the query path from the data tree.
		\item Read and update the data result log.
	\end{enumerate}
\end{itemize}

Observe that eviction Steps 1 and 2 perform only read accesses and do not
conflict with the query protocol.
Eviction Step 3 however updates the data tree and the stash and requires 
synchronization to avoid inconsistencies.

\paragraph{Insight 1: Separate Trees for Queries and Evictions}   
One way to to synchronize this is to {\em perform queries on a read-only copy of
	these data structures while the data tree and stash updates during evictions
	happen on a writable copy never accessed by the actual queries}.  Once
eviction completes updating the writable copies,
their contents can be (efficiently and securely) copied (by the server) to
the read-only data tree and stash version, and made available for future
queries. This is aided by two server-hosted data structures: 

\begin{itemize}
	\item {\em Write-only tree}: A write-only tree (``W/O tree'') is
	initialized with the same contents as the read-only data tree.  The
	W/O tree is updated during evictions while the data tree is used to
	satisfy queries in the background.  This is possible because queries
	do not update the data tree and queried blocks that are updated with
	new data make it back to the data tree only during evictions.
	
	\item {\em Temporary stash:} During the eviction of blocks along a
	specific path of the write-only tree (eviction path), blocks that cannot be
	accommodated are placed in a ``temporary stash'' (not accessible to
	queries) on the server.
	
	Once the entire eviction path is updated, the eviction path is
	copied from the write-only tree to the data tree, and the temporary stash
	is made available for queries by efficiently replacing (e.g., by a
	simple reference swap) the main stash with the temporary
	stash.  Note that at this stage the contents of the main stash have
	already been evicted to the data tree and the temporary stash, and
	thus can be replaced without losing track of data.
\end{itemize}

\paragraph{Insight 2: Multi-phase Evictions}  
To execute evictions without blocking queries, \sysname~ splits evictions
into

\begin{itemize}
	\item {\em Processing Phase}
	
	\begin{enumerate}
		\item Fetch eviction path buckets from the {\em write-only} tree 
		\item Fetch current stash
		\item Evict contents of the stash and the eviction path
		\item Create a new path and temporary stash locally 
		\item Write back the updated path to the {\em write-only} tree
		\item Write back the temporary stash to the server
	\end{enumerate}
	
	
	\item {\em Commit Phase}
	%
	
	\begin{enumerate}
		\item Update the position map for the blocks that have been evicted to the {\em write-only} tree
		\item Copy the eviction path from the {\em write-only} tree to the data tree (server-side copy)
		\item Swap reference of the main stash (previously used to satisfy queries) with the temporary stash 
		\item Clear the query log 
		\item Clear the data result log
	\end{enumerate}
	
\end{itemize}

An eviction can perform the processing phase in its entirety before
performing the commit.  Eviction processing -- which is significantly more
expensive than the commit -- can be executed in parallel with queries.  This
allows \sysname~ to block queries only while an eviction {\em commits}.

\paragraph{Insight 3: An Oblivious Data Structure for
	Storing and Privately Accessing Query Results}
Since queries do not need to wait for evictions, multiple query rounds may
run by the time one eviction completes.  Blocks accessed in these query
rounds need to be stored somewhere until an eviction subsequently replaces them on the tree. Each round of parallel queries that is executed in the
background while an eviction takes place, generates a {\em data result log}
(DRL) containing the blocks that have been accessed by its queries.

These DRLs need to be maintained separately on the server until their contents can
be evicted back into the data tree (which can then be accessed by future
queries). Further, reading all such DRLs in their entirety in the query protocol (Step 5 in Figure \ref{query_overview}) may
be too expensive.  Instead we propose a mechanism to efficiently query for
particular items from the DRLs without leaking their identity to the server.

To this end, DRLs that are pending evictions are stored in an oblivious data
structure, namely the {\em DR-LogSet}. This allows {\em clients to efficiently query for
	particular blocks from the DRLs without leaking to the server: (i) the identity of the
	block, and (ii) the block's last access time.}

\subsection{Parallel Evictions}
\label{overview:parallel_evictions}

\begin{figure}[t]
	\centering
	\includegraphics[scale=0.10]{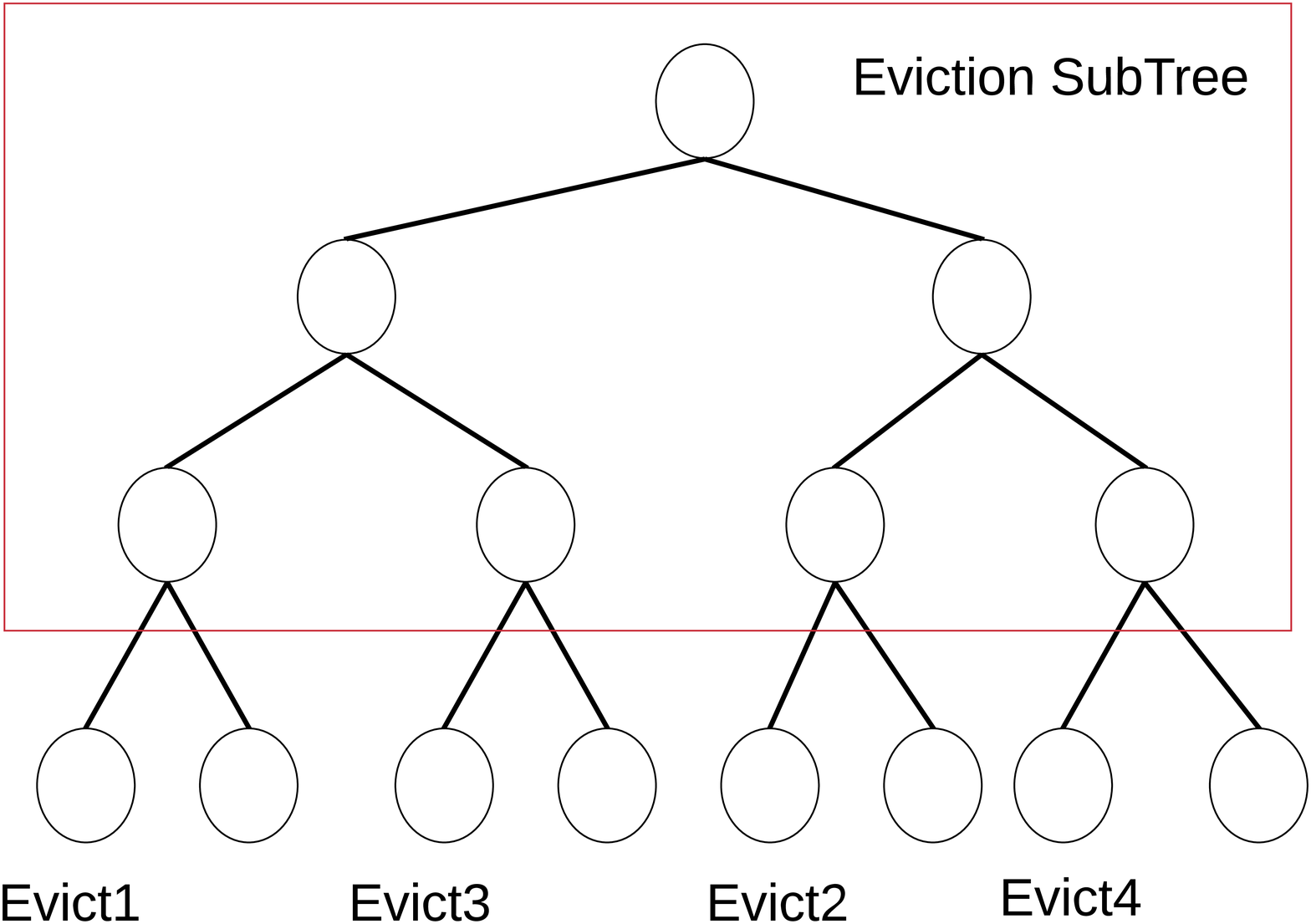}
	\vspace{-0.5cm}
	\footnotesize{\caption{\footnotesize Eviction subtree (EST) defined by 4 parallel eviction clients evicting to paths in reverse lexicographical order of their leaf IDs.\label{est}}}
	\vspace{-8pt}
\end{figure}

One challenge here is that if blocks
are evicted to the data tree from a single DRL at a time, the DR-LogSet may grow
uncontrollably -- by the time a DRL is cleared, several new DRLs will have
been created.

To ensure that the DR-LogSet remains bounded to an acceptable size, evictions
must start as soon as a DRL has been added, even if a previous eviction has
not finished.  Effectively, evictions must execute in parallel.  One obvious
roadblock here is that multiple evictions cannot commit simultaneously
because during the commit phase, the same data structures need to be
updated. As we will see next, a viable, efficient approach is to perform the (expensive)
processing phases in parallel while serializing the commits.

\begin{figure}[t]
	\centering
	\includegraphics[scale=0.025]{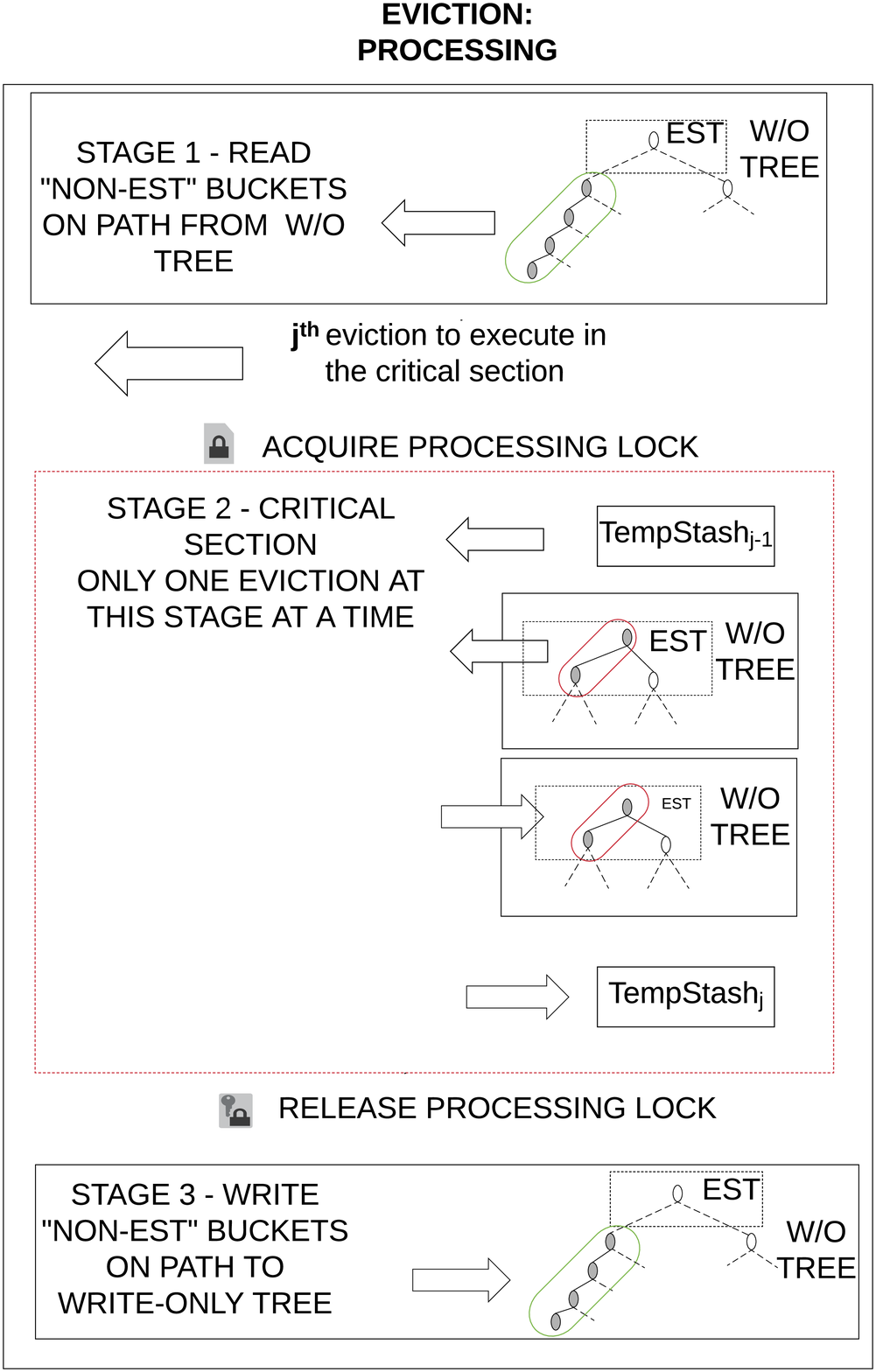}
	\vspace{-1.5cm}
	\caption{\footnotesize Eviction processing is divided into three
		stages.  Fixing the maximum number of evictions that can be executed in parallel 
		at initialization allows \sysname~ to define a maximal sub-tree outside of which 
		all ongoing evictions cannot overlap, namely the EST.
		Stages 1 and 3 read and write back the non-EST buckets from eviction path,
		respectively. An eviction executing Stage 1 does not overlap
		with evictions in Stage 3. Stage 2 is the
		critical section which updates the EST and is accessed after acquiring the processing
		lock.  The execution of the critical section defines an ordering of
		evictions, which is used later to serialize
		commits.\label{evict_process}}	
	\vspace{-8pt}
\end{figure}

\paragraph{Insight 4: Identifying Critical Sections for Parallel Eviction Processing}
Facilitating parallel eviction processing is challenging due to overlapping
accesses to the {\em write-only} tree in Step 5 of the processing phase: two
independent paths being evicted to in parallel will invariably intersect at
some level of the write-only tree.  Updates to any buckets residing on the
paths' intersection need to be synchronized.
A key insight here is that we can precisely predict the overlaps! This is
because evictions are performed to deterministically selected paths.

First, recall from Sections \ref{overview:preliminaries} and
\ref{overview:parallel_evictions} that evictions are performed to data tree
paths in reverse lexicographical order of their corresponding leaf IDs.
Since each path is represented by a {\em unique} lexicographical
representation, {\em any $k$ consecutive eviction paths are deterministic}. 
Further, if only these $k$ (determined based on system load) evictions are allowed to execute in parallel,
then the overlaps between the paths can be predicted precisely.  As a
result, we can {\em define a maximal subtree outside of which {\em any} $k$
	consecutive eviction paths will never overlap}.

\begin{itemize}
	
	\item The {\em eviction subtree} (EST) is a subtree of the write-only tree
	containing the root and the buckets overlapping between any $k$ paths
	corresponding to consecutive evictions allowed to execute in parallel.  For
	$k$ consecutive evictions executing in parallel, the height of the EST is $h = \log{k+1}$ (Section \ref{desc:est}).  Figure \ref{est} shows an example EST with 4
	parallel evictions.

	\item {\em Fine-grained locking for eviction subtree access.~}  Accesses to
	the eviction subtree constitute the critical section of the {\em processing phase} and is protected by a mutex, namely the {\em processing lock}.  These accesses must be performed {\em atomically}
	by a single eviction client at a time.  Multiple evictions can execute the rest of the steps of
	the processing phase in parallel.
	
\end{itemize}



\begin{figure}[t!]
	\includegraphics[scale=0.03]{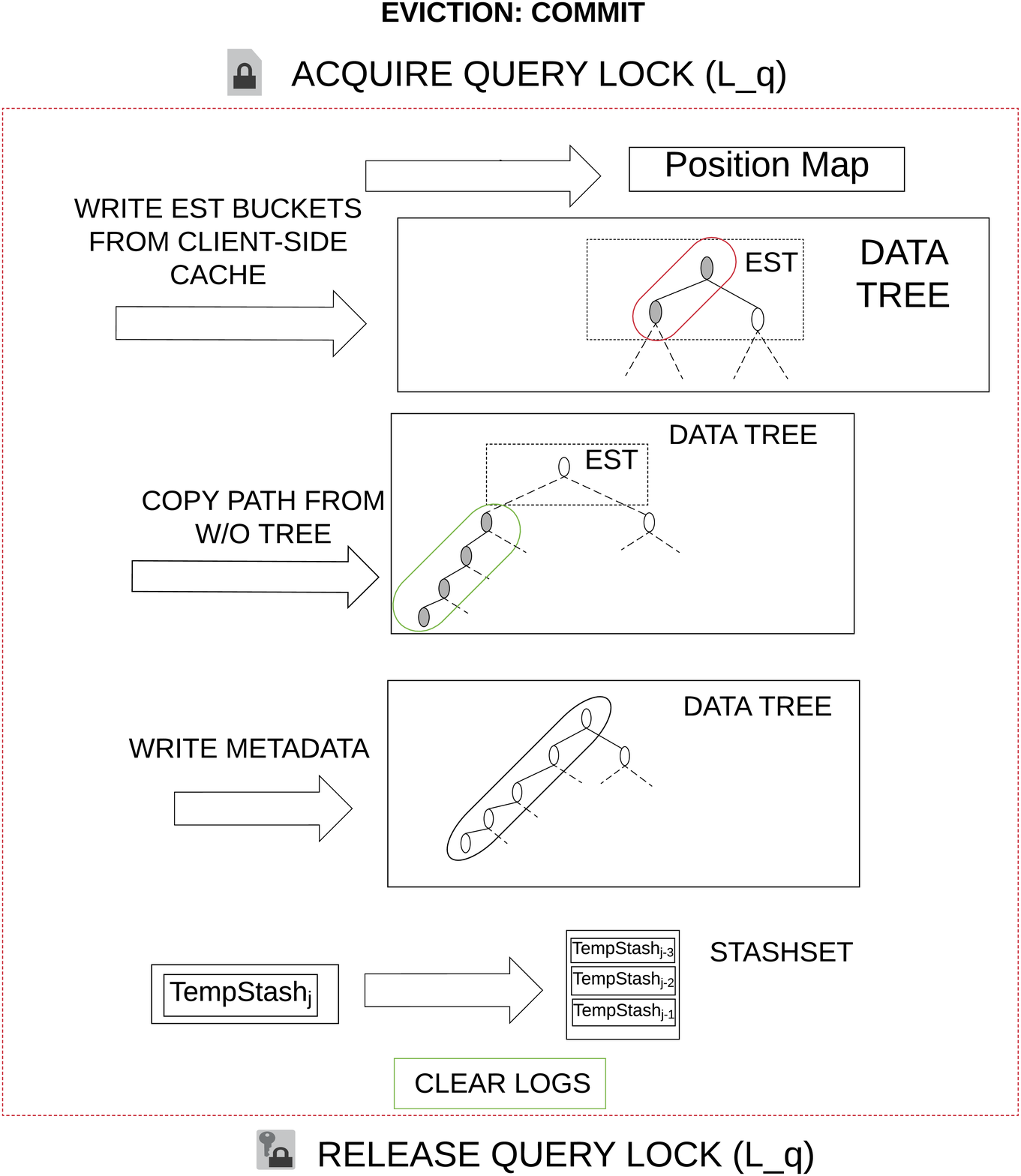}
	\vspace{-0.6cm}
	\caption{\footnotesize Eviction commit includes updating the position map, updating the
		data tree path and adding the temporary	stash to the StashSet. Only one eviction can commit 
		at a time, while also blocking queries during commit.}\label{eviction_commit}
	\vspace{-5pt}
\end{figure}

\paragraph{Insight 5: Asynchronous Commits}
Updates performed in the critical section of the processing phase determine
the behavior of future evictions and queries.  An eviction that enters the
critical section uses results generated by previous evictions.  Thus, {\em
	if commits are serialized based on the order in which evictions enter the
	critical section, additional synchronization is not necessary} to ensure
consistency of data structures. However such fully serialized commits may result in unnecessary and
prolonged waits.

To further increase parallelism, \sysname~ allow asynchronous eviction
commits.  Concurrent evictions can commit in any order.  To enable this,
information from each eviction is securely persisted server-side and
reconciled later when all preceding clients in the serial
order also finish their commits.  This is facilitated using an 
oblivious data structure: 

\begin{itemize}
	
	\item The {\em StashSet} is similar in design and functionality to
	the DR-LogSet.  In particular, the StashSet stores the temporary
	stashes for out-of-order commits.  Critically, the StashSet also
	allows oblivious queries -- {\em clients can efficiently query for
		particular blocks in the StashSet without leaking to the server: (i)
		the identity of the block, and (ii) the last access time of the
		block.}.
	
\end{itemize}	


\section{Technical Description}
\label{desc}
%

\paragraph{Notation} 
The server stores $N$ blocks of data, with logical address in $[0, N-1]$.  A {\em real data} block with logical address $id$ is denoted by $b_{id}$. Dummy blocks are assigned addresses (just for reference) outside the address space of real data blocks. A dummy block with address $i$ is denoted as $d_{i}$. Once a data block is retrieved from the server, it is uploaded back only after re-encryption with fresh randomness.

\paragraph{Temporary Identifiers}
Prior to execution, both queries and evictions are assigned {\em temporary identifiers}, which are extensively used for synchronization without inter-client communication (as will be discussed later). Specifically

\begin{itemize}
	\item {\em Query identifier: }  Each query in a round is logically identified by a unique {\em query identifier} $ 0 \leq i \leq c-1$. The query identifier reflects the order in which the queries start execution by appending an entry to the query log. Clients learn their query identifiers prior to query execution from the query log.

	\item {\em Eviction identifier.~} \sysname~ synchronizes evictions based on the critical section. Specifically, the order in which evictions execute the critical section, is also used for ordering the commits, for the sake of consistency. 
	
	For this, a {\em processing counter}, is stored server-side to track the number of evictions that have executed the critical section since initialization, .  While in the critical section, an eviction reads the value of this processing counter, which becomes its {\em temporary eviction identifier}.  Before exiting the critical section, the processing counter is incremented. 
 \end{itemize}

\subsection{DR-LogSet}	
\label{desc:drlogset}
The DR-LogSet is used to store and privately query DRLs generated by previous query rounds until an eviction can write back contents to the data tree. The DR-LogSet includes the current DRL (of size $c$), and $k \leq c$ DRLs generated by rounds of queries that finished execution while an eviction was being performed in the background.

\begin{itemize}
	\item {\em Bigentry logs.~}  Except for the current DRL, all other
previously generated DRLs are stored in the DR-LogSet as ``bigentry logs'' of size $2\cdot c$ -- each bigentry log is composed of a random permutation of the (re-encrypted) blocks of a previously generated DRL combined with an
additional $c$ random dummy blocks. Dummy blocks are assigned identifiers from 0 to $c-1$.

	\item {\em Temporal ordering \& log identifiers.~} Bigentry logs are ordered by ascending insertion time and assigned identifiers, $l_{0}, l_{1},
\ldots, l_{k-1}$, with $l_{k-1}$ being the ID of the log inserted most recently.

	\item{\em Search index:} A search index is appended to 
	each bigentry log to allow retrieval of specific blocks efficiently. This is simply an encrypted list of block IDs ordered by their corresponding positions in the bigentry logs. Due to small-sized entries, the list is small and can be downloaded entirely  per access to determine the location of real/dummy blocks in a particular bigentry log.

\end{itemize}

	\begin{algorithm}[t]
		\caption{$\mathsf{readLogSet(id)}$}\label{readlogset}
		\footnotesize
		\begin{algorithmic}[1]
			\State Read current DRL
			\State $i \gets$ Current DRL size
			\For{$j \in [k, k-1, \ldots, 1]$}
			\State $index\_blk \gets$ Read search index for $l_j$
			\If {$id \in index\_blk$}
			\If{$id \notin DRL$}
			\State Read $b_{id}$ from $l_{j}$
			\State Remove $id$ from $index\_blk$
			\Else
			\State Read dummy block $d_i$ from $l_{j}$ 
			\State Remove $b_{id}$ from $index\_blk$
			\EndIf
			\Else
			\State Read the $d_i$ dummy block from $l_{j}$
			\EndIf 
			\State Reencrypt and write back $index\_blk$
			\EndFor
		\end{algorithmic}
	\end{algorithm}

\begin{small}
	\begin{algorithm}[t]
		\caption{$\mathsf{writeLogSet(blk, i)}$}\label{writelogset}
		\footnotesize
		\begin{algorithmic}[1]
			\State $blk \gets$  Queried block to be appended to DRL
			
			\State Append encrypted $blk$ to current DRL
			\State Read $l_{i}$ from DR-LogSet
			\State $l_{i}$.reshuffle(rand)
			\State Write back $l_{i}$ to temporary workspace
			
			\If{$i = c-1$}
			\If{less than $c$ bigentry logs in DR-LogSet}
			\item[] \begin{center} (Create new bigentry log) \end{center}
			\State $j \gets$ Number of logs in the DR-LogSet 
			\State Initialize $l_{j+1}$ with size $2c$ blocks
			\State $l_{j+1} = blk + drl_{curr} + dummy$
			\State $l_{j+1}.reshuffle(rand)$
			\State  Append search index to $l_{j+1}$
			\State  Reencrypt and write back $l_{j+1}$ to the DR-LogSet
			\State  Initialize empty DRL for new round of queries
			\Else
			\item[] \begin{center}(Wait until next eviction commit)\end{center}
			\EndIf	
			\EndIf 
		\end{algorithmic}
	\end{algorithm}
\end{small}

\paragraph{Querying the DR-LogSet}
The $\mathsf{readLogSet(id)}$ (Algorithm \ref{readlogset}) protocol takes as input the logical ID $id$ of the block being queried and performs the following steps
%
%
%

\begin{enumerate}
	\item Read the current DRL.
	\item {\em Privately query bigentry logs:} Read {\em one} block each	from the bigentry logs in {\em descending order} of log ID -- recent logs before old ones.  
	\begin{enumerate}
		\item If the queried block ID $id$ is present in a
		bigentry log as determined from the corresponding search index.
		and the block is not present in the current DRL, it is retrieved from the
		bigentry log and ID $id$ is removed from its search index.. 
		\item If the block is also present in the current DRL, a dummy
		block $d_i$ is read from the bigentry log and $id$ is removed from the search index.
		\item If the queried block is not in the bigentry log, a dummy
		block $d_i$ is read. 
	\end{enumerate}
\end{enumerate}

\paragraph{Periodic Shuffling}
 The intuition behind the query protocol is to ensure that the server does not learn the identity of the block being queried.
Since each of the bigentry logs contains $c$ dummy blocks, they need to be shuffled once every $c$ accesses thus ensuring that it does not run out
of {\em unique} dummy blocks. This is performed efficiently in the background as part of an update 
which reshuffles and writes-back a bigentry log to a temporary {\em write-only workspace} on the server. {\em The reshuffled bigentry logs in the workspace replace the old versions in the DR-LogSet after a round of $c$ queries (as part of the query protocol, Section \ref{desc:query}). }

\paragraph{Updating the DR-LogSet}
The $\mathsf{writeLogSet(blk)}$ (Algorithm \ref{writelogset}) protocol takes as input: i) the block $blk$ that needs to be appended to the current DRL, and ii) the client query identifier $i$, and performs the following steps

\begin{enumerate}
	\item Append $blk$ to the current DRL
	\item {\em Reshuffle specific bigentry log:} Each bigentry log is uniquely identified by its ID $l_{i}, 0 \leq i < c$. Each client registering a new query in the query log is also uniquely identified in the current query round by its query identifier. The client with query identifier $i$ reshuffles the bigentry log identified by ID $l_i$ 
	​
	\item {\em Create new bigentry log:} If this update is performed as the last
	 query in a round of $c$ queries, the client initializes a new ``bigentry log'' with $2\cdot c$ blocks.  The block queried by the client and the up-to-date blocks from the current DRL are added to the newly created bigentry log while filling up the remaining part with dummy blocks.

\end{enumerate}

Data blocks of the DR-LogSet bigentry logs are eventually evicted to the data
tree by future evictions.  Specifically, once the current eviction
completes, the next eviction will evict blocks from the ``oldest'' bigentry log. Therefore, 
if the DR-LogSet already contains $c$ bigentry logs, a new log cannot be added until at least one eviction is completed.

\paragraph{Obliviousness} 
The DR-LogSet query protocol (Algorithm \ref{readlogset}) ensures that the server does not learn: i) the identity of the block being queried, and ii) the last access time of the block. 

\begin{itemize}
	\item A block is read from each of the bigentry logs, in a {\em specific order}, regardless of the  bigentry log that actually contains the block. Due to the random permutation of blocks in each bigentry log, the block read from each log appears random to the server.
	\item Blocks that have been read once from a bigentry log, have their corresponding entries removed from the search index. Indices are updated client-side and encrypted with semantic security, preventing the server from learning which entry has been removed. As a result, the same block is not read from the same bigentry log ever again. 
	\item Due to the round robin reshuffling,, a bigentry log is accessed $c$ times
	before it is replaced by an independently reshuffled version of the log (unless the bigentry log is cleared by an eviction before).  As each log
	contains $c$ dummy blocks, a different dummy block can be read for
	each of the $c$ accesses before the next reshuffle. 
\end{itemize}	
	

\begin{customthm}{1}
	\label{drl_lemma}
	The accesses to the DR-LogSet produce transcripts that are indistinguishable
	and independent of the block that is being queried.
\end{customthm}

\begin{proof}
	The DR-LogSet is accessed through $writeLogSet(blk, i)$ and $readLogSet(id)$ where $id$ is the ID of the block being queried 
	and $blk$ is the queried data block. We show that the transcripts generated by these operations 
	are independent of the inputs  -- thus the server does not learn $blk$ or $id$ by observing 
	the corresponding transcripts. To achieve this, we prove the existence a simulator with access to 
	only public information that can generate indistinguishable transcripts to the ones generated by these
	operations. 
	
	\begin{itemize}
		\item { $\Access^{Real}(writeLogSet(blk, i))$:} The real access transcript produced by $writeLogSet(blk, i)$:

		\begin{enumerate}
			\item Append an encrypted block to the DRL
			\item Randomly shuffle bigentry log $l_{i}$ from the DR-LogSet where $i$ is the query identifier. Write the reshuffled log back to the {\em temporary workspace}.
			\item If the current DRL is full, add a new log of size $2 \cdot c$ 
			to the DR-LogSet containing blocks 
			from the DRL and dummy blocks, re-encrypted and randomly shuffled together.
		\end{enumerate}
		
		\item { $\Access^{Sim}(writeLogSet(blk,i))$:} Transcript produced by a simulator with only public knowledge of the query identifier, which depends only on the number of queries that have executed in the current round:
		
		\begin{enumerate}
			\item Append a block with random contents to the DRL. 
			\item Simulate reshuffling bigentry log $l_i$ by creating 
			a new log $l_{i}$ with only random blocks, and write it to the {\em temporary workspace}. 
			\item If the current DRL is full, add a new log of size $2 \cdot c$ 
			to the DR-LogSet containing random blocks.
		\end{enumerate}
		
	\end{itemize}
	Due to semantic security,
	\begin{small}$$\Access^{Real}(writeLogSet(blk, i)) \approx_c { \Access^{Sim}(writeLogSet(blk, i))}$$\end{small}
	
	Next, we show the existence of a simulator with only public information
	that generates access transcripts indistinguishable 
	to the access transcripts generated by $readLogSet(id)$.
	%

	\begin{itemize}
		\item { $\Access^{Real}(readLogSet(id))$:} The real access transcript produced by $readLogSet(id)$ -- 
		
		\begin{enumerate}
			\item Read DRL
			\item Read the index block for each log in the DR-LogSet
			\item Read {\em one} block from each log in the DR-LogSet
			\item Update and write back the index block for each log in the DR-LogSet.
		\end{enumerate}
		
		\item { $\Access^{Sim}(readLogSet(id))$:} Transcript produced by a simulator with only knowledge  of the blocks accessed since a log $l_i$ has been previously reshuffled --
		
		\begin{enumerate}
			\item Read DRL
			\item Read the index block for each log in the DR-LogSet
			\item For each log $l_i$, if $j$ accesses have been performed since the last reshuffle, 
			read a random block from the log that hasn't been read in the last $j$ accesses.
			\item Reencrypt and write back the index block for each level.
		\end{enumerate}
	\end{itemize}
	
	Due to semantic security of the encrytion scheme ,
	$$\Access^{Real}(readLogSet(id)) \approx_c { \Access^{Sim}(readLogSet(id))}$$

\end{proof}

\subsection{Eviction Subtree }
\label{desc:est}
%


\paragraph{Reverse Lexicographical Ordering of Leaf Identifiers}
Recall from Section~\ref{overview} that evictions are performed to paths in the data tree in reverse lexicographical order of their corresponding leaf IDs. Intuitively this results in all $N$ paths of the tree being selected once before the same path is selected again for eviction.  \sysname~ stores a global counter, $ctr$, tracking the number of evictions that have been executed since initialization, and the {\em next} $k$ successive evictions are to paths that have
reverse lexicographical representations matching the binary representations 
of $v = (ctr+1) \bmod N, \ldots (ctr+k) \bmod N $ respectively. Let these paths be $p_1, p_2, \ldots, p_k$.

Since each path is represented by a {\em unique} lexicographical
representation, any $k$ consecutive eviction paths are deterministic.  Further, if only these $k$ evictions are allowed to execute in parallel, then the
overlaps between the paths can be predicted precisely.  As we
show next, if the maximum number of consecutive evictions that can be executed in
parallel is fixed at initialization, to say $k$, we can define a maximal subtree outside of which the $k$ successive eviction paths will never overlap in the write-only tree. 
We first present a related result.

\begin{customthm}{2}
	\label{thm:rev_lex}
	The length of the longest common suffix between the binary representations of any
	$k \leq N/2$ consecutive integers selected from the integer modulo group of
	$N$, i.e.  $\{0,1, \ldots, N-1\}$, is bounded by $\log{k}$.
	
\end{customthm}

\begin{proof}
	We prove this by contradiction. W.l.o.g let $x_1, x_{2}, \ldots x_{k}$ be $k \leq N/2$  {\em consecutive} integers in $\mathbb{Z}$. Then, for any $x_i$ and $x_j$, where $i, j \in (1,2,...k)$ and $x_j > x_i$,
	$x_j - x_i \leq k$. 
	

	
	Also, then $x_1 \bmod N, x_{2} \bmod N, \ldots x_{k} \bmod
	N$ are $k$ consecutive integers from the integer modulo group of N.  First,
	we show that $x_j \bmod N - x_i \bmod N \leq k$.
	
	W.l.o.g. we can write $x_i = Np_1 + r_1$ and $x_j = Np_2 + r_2$, where $p_1, p_2, r_1, r_2$ are positive integers and 
	$r_1, r_2 < N$. In other words, $r_1 = x_i \bmod N$ and $r_2 = x_j \bmod N$.
	
	Then 
	
	\begin{center} 
		$x_j - x_i \leq k$ \\
		$\Leftrightarrow$ \\
		$N(p_2 - p_1) + r_2-r_1 \leq k$
	\end{center}

	Now, if $p_2 < p_1$, $p_1 - p_2 \geq 1$. This implies $x_i - x_j = N(p_1 - p_2) + (r_1 - r_2) \geq N + (r_1 - r_2) = N - (r_2 - r_1) $. But $(r_2 - r_1) \leq N-1$, since  $0 \leq r_1, r_2 < N$. This would imply that $ x_i - x_j \geq N - (N - 1) = 1$ which contradicts the fact that $x_j > x_i$. Thus, 
	$p_2 \geq p_1 \Rightarrow (p_2- p_1) \geq 0$.

	
	Thus  	
	\begin{center}
		$r_2-r_1 \leq k$ \\
		
		$\Leftrightarrow$ \\
		
		\begin{equation}
		x_j \bmod N - x_i \bmod N \leq k 
		\end{equation}
	\end{center}
	
	Next, consider that $b_i$ and $b_j$ are strings (of length $\log{N}$), corresponding to the 
	binary representations of $x_i \bmod N$ and $x_j \bmod N$ respectively. Let us assume that the length of the longest common suffix between $b_i$ and $b_j$ is $\log{k+1}$. Since $k \leq N/2$, $\log{k+1} \leq \log{N}$. By the definition of the longest common suffix, $b_i$ and $b_j$ must necessarily differ in their $(logk+1)^{th}$ bit. In that case, $b_j - b_i$ will be of the form ($\ldots, 1_{k+1}, 0_{k},0_{k-1}, \ldots$). In other words, the $(logk+1)^{th}$ bit in the binary representation of $x_j \bmod N - x_i \bmod N$ will be 1. But this implies that $x_j \bmod N - x_i \bmod N \geq 2^{logk+1} = 2k$ which contradicts 
	Equation 1.
\end{proof}



By  Theorem \ref{thm:rev_lex}, given any two
	values within the next $k$ consecutive values of $v$, say $v_1$ and $v_2$, the binary representations of $v_1$ and
	$v_2$ cannot have a common suffix of length greater than $\log{k}$. On further introspection, it can be observed that the length of the longest 
		common suffix between $v_1, v_2, \ldots, v_k$, corresponds to the levels of the tree  where the correspondingly chosen $k$ eviction paths, $p_1, p_2, \ldots p_k$ can possibly intersect.
	
{\em In effect, $k$ parallel evictions to paths determined by the reverse-lexicographical ordering of leaf IDs that started in
succession, can overlap with each other on at most the first $\log{k}+1$ levels of the
tree. }

%

%

\begin{definition}
	The eviction subtree (EST) is a full binary tree of height $h = \log{k+1}$, 
	where $k$ is the maximum number of consecutive evictions that are allowed to execute in parallel at any given time,
	such that the root of the write-only tree is also the root of the eviction subtree	
	
\end{definition}

\noindent
{\em Critical section:} Importantly, while writing back to the write-only tree,  updating the
EST and uploading the temporary stash is the critical section and is
performed atomically by only one eviction at a time, enforced by a mutex (processing lock).  The rest of the path
can be updated asynchronously.

\paragraph{Fixing the Number of Parallel Evictions} 
\sysname~ fixes the maximum number of {\em consecutive} parallel evictions that can be
executed at a time, $k \leq N/2$, during initialization. For e.g., if $e_1, e_2, \ldots, e_k$ are the $k$ consecutive evictions that are executing currently, then, eviction $e_{k+1}$ cannot start execution until eviction $e_1$ completes
The value of $k$ will usually be determined by the system load and in general can be set equal to the query log/DRL size ($c$). 
 Without inter-client communication, one way to achieve this is by maintaining a server-side log of all ongoing evictions. Specifically,

\begin{itemize}
	\item {\em Eviction log.~} The eviction log stores information about all currently ongoing evictions. Prior to execution, an eviction client reads the eviction log and appends an entry, only if its eviction path does not overlap outside the write-only tree with any of the ongoing evictions. 	When an eviction commits, the entry from the eviction log is removed. 
	 Accesses to the log are synchronized using a mutex, namely the {\em eviction lock}. It may be evident that the eviction log performs the same role as the query log. 
\end{itemize}

\subsection{Parallel Eviction Processing}
\label{desc:evict_processing}
\begin{small}
	\begin{algorithm}[t]
		\footnotesize
		\caption{$\mathsf{Evict.Process}$}\label{evict_process_algo}
		\begin{algorithmic}[1]
			\State $ctr \gets$ eviction counter
			\item[] // Update eviction log 
			\State Read eviction log
			\If{$(ctr-k) \notin$ eviction log}
			\State $EvictionLock.lock$
			\State Append $ctr$ to eviction log
			\State $EvictionLock.Unlock$
			\Else
			\State Wait for eviction $ctr-k$ to finish
			\EndIf
			\item[] // Processing Stage 1
			\State $path  \gets WriteOnlyTree.readPath(ctr)$
			\item[] // Processing Stage 2 (Critical section)
			\State $ProcessingLock.lock$
		
			\State $i \gets$ Temporary eviction ID
			\State  Read  $TempStash_{i-1}$
			\State $BucketsFromEST \gets$ Read Buckets from EST that intersect with $path$
			\State $path.UpdateBuckets(BucketsFromEST)$
			\State $union = TempStash_{i-1} + DRL + path$
			\State $path =  union.EvictToPath$
			\State $TempStash_{i} \gets$  Blocks left in $union + dummy$ 
			\State Write back $TempStash_{i}$
		
			\For{ $bkt \in path.Buckets$}
			\If{$bkt \in EST$}
			\State {Write $bkt$ to write-only tree}
			\EndIf
		
			\EndFor
		
			\State $ProcessingLock.Unlock$
			\item[] // Processing Stage 3				
		
			\State $WriteOnlyTree.write(path, ctr)$
		\end{algorithmic}
	\end{algorithm}
\end{small}

\begin{small}
	\begin{algorithm}[t]
		\footnotesize
		\caption{$\mathsf{Evict.SyncCommit(i)}$}\label{evict_sync_commit_algo}
		\begin{algorithmic}[1]
			\State $QueryLock.lock$
			\State Update position map and metadata on data tree path
			\State Copy EST buckets from client-side cache to data tree
			\State Copy remaining buckets on eviction path from write-only tree to data tree
			\State $Stash = TempStash_{i}$
			\State Clear query log 
			\State Clear bigentry log from DR-LogSet
			\State $QueryLock.Unlock$
		\end{algorithmic}
	\end{algorithm}
\end{small}

\paragraph{Processing Protocol}
The parallel eviction processing protocol (Algorithm \ref{evict_process_algo}) includes the following steps

\begin{enumerate}
	
	\item {\em Stage 1} -- Read the non-EST buckets on the eviction path from
	the write-only tree.
	
	\item {\em Stage 2} -- A critical section which requires acquiring the
	{\em processing lock}. It includes the following substeps
	
	\begin{enumerate}
		
		\item Read the temporary stash uploaded by the eviction that {\em
			last} executed the critical section, denoted $TempStash_{i-1}$,  separately maintained on the
		server.
		
		\item Read buckets in the EST along the eviction path from the write-only tree.
		
		\item Write back updated EST buckets along the eviction path, and the new  temporary stash, $TempStash_i$.
		
	\end{enumerate}
	
	\item {\em Stage 3} -- Write back non-EST buckets on the eviction path to the write-only tree.
	
\end{enumerate}

\paragraph{Processing Cost} 

\begin{itemize}
	\item {\em Non-blocking stages:} Stage 1 and Stage 3 read and write back $\O(\log{N})$ buckets along 
	the eviction path (excluding the small number of EST buckets) on the 
	write-only tree. As discussed in Section \ref{overview:preliminaries}, 
	buckets contain $\O(\log N)$ blocks. Thus, Stage 1 and 3 both feature an 
	asymptotic access complexity of $\O(\log^{2} N)$ blocks. Note that these steps can be performed in
	parallel without blocking queries and other ongoing evictions.


	\item{\em Critical section:} Only {\em one} eviction can execute in stage 2 at a time.  This includes
	reading and writing back buckets along a path from the eviction subtree. The height of the eviction subtree is $\log k + 1$ (Section \ref{desc:est}).
	The overall asymptotic access complexity of this stage is $\O((\log{k}+1)\cdot \log{N})$ blocks. 
	This is significantly less expensive than Stage 1 and 2 for
	realistic deployment scenarios.  In fact, stage 2 becomes 
	expensive only when $N$ parallel eviction are allowed to execute in parallel!

\end{itemize}

\subsection{Synchronous Commits}
\label{desc:sync_commits}
Before describing the more complex asynchronous commit mechanism, we present a relatively simple design for {\em synchronous commits -- evictions commit in the order in which they execute the critical section}.
The challenge here is to persist eviction-specific changes to the eviction subtree. Specifically 
after an eviction writes to the eviction subtree, subsequent evictions that execute in the critical section can overwrite these contents before the eviction commits.  If the contents of the eviction subtree are directly copied to the data tree as part of the commit, this may lead to inconsistencies.

One possible solution is to locally cache the changes to the eviction subtree in a {\em client-side cache}, and use this to update the contents of the data tree during commits. Specifically, the {\em client-side cache} includes: i) buckets in the eviction subtree written in Stage 2 of the processing protocol, and (ii) the temporary stash created during eviction.

\paragraph{Synchronous Commit Protocol}
The synchronous commit protocol (Algorithm \ref{evict_sync_commit_algo}) uses the {\em temporary eviction identifier}, $i$, of the eviction and  performs the following ordered steps 

\begin{enumerate}
	\item  Update the position map and eviction path metadata. 
	\item Copy contents of the eviction subtree buckets from the client-side cache to the data tree.
	\item Copy remaining eviction path from the write-only tree to the data tree.
	\item Set the temporary stash from the client-side cache as the stash of the data tree.
	\item Clear the query log and the bigentry log corresponding to the eviction from DR-LogSet.
\end{enumerate} 

\paragraph{Updating Bucket Metadata} 
Metadata on the data tree path is 
updated during a commit, while accounting for the current state of the bigentry logs in the DR-LogSet.  Specifically, as blocks from a bigentry log are evicted to the write-only tree during eviction processing, some blocks in the log may be accessed by queries executing in the background.  Since, there are more recent copies of these blocks in other bigentry logs, the old copies should not be made available for queries. Subsequent queries for these blocks can access the upto-to-date copies from the more recently created bigentry logs. 

Observe that these blocks will already be on the eviction path in the write-only tree by the time an eviction commits. 
As a result, these blocks will also be included when the contents of the eviction path are copied from the write-only tree to the data tree during the commit. 

{\em Instead of explicitly removing these blocks by re-scanning the entire eviction path, which will certainly be expensive,  \sysname~ indirectly invalidates these blocks by not updating their corresponding metadata and position map entries on the data tree path. Consequently, old copies are inaccessible to queries (which reads the position map first to locate a particular block) and are removed from the eviction path during later evictions (using the metadata entries on the path)}.

Finally, recall that once a block has been accessed from a bigentry log, its corresponding entry is removed from the search index. The search index allows \sysname~ to identify blocks that have been accessed from the bigentry log while the eviction was executing.

\subsection{StashSet}
\label{desc:stashset}


\paragraph{Storing and Privately Querying Temporary Stashes} 
With synchronous commits (Section~\ref{desc:sync_commits}), the temporary
stash created by an eviction can straightforwardly replace the main stash
and satisfy future queries after the commit.  However, this is not the case when evictions commit asynchronously. 

For example, consider an ordering of evictions established by the execution
of the critical section, $e_{i-1} < e_{i} < e_{i+1}$.  Also, let $e_{i+1}$
commit when $e_{i-1}$ has committed but $e_{i}$ is yet to commit.  In this
case, contents that are evicted to a path, $p_{i}$ by $e_{i}$ from the
temporary stash created by $e_{i-1}$ (denoted by $TempStash_{i-1}$), will
not be available for queries until $p_{i}$ is updated in the data tree. 
Thus, replacing $TempStash_{i-1}$ with $TempStash_{i+1}$ as the main stash
will lead to data loss.

\paragraph{StashSet}
To overcome this, instead of storing a single stash, \sysname~ stores a set
of {\em temporary stashes} which were created and uploaded by evictions that have committed asynchronously before previous evictions could be completed. The StashSet is structurally similar to the DR-LogSet and contains $k \leq c$ temporary stashes, organized as follows

\begin{itemize}
	
	\item {\em Temporary stashes:} 	Each temporary stash contains upto $\mathsf{MaxStashSize}$ real blocks and at least $c$ dummy blocks, where $\mathsf{MaxStashSize}$ is the stash size determined according to \cite{ringoram}. Blocks are encrypted and randomly shuffled.
	
	\item {\em Search index:} A search index (list of block IDs) tracks the location of real and dummy blocks. The entire (small-sized) search index is downloaded per access.

	\item {\em Temporary stash identifier:} The temporary stashes in the StashSet are identified by the {\em temporary eviction identifiers} of evictions that created the stashes. For example, if 
	the eviction with temporary eviction identifier $k$ created a temporary stash currently in the StashSet, then the temporary stash is identified as $TempStash_{k}$

\end{itemize}

Each temporary stash is periodically reshuffled (exactly once every $c$ accesses) to ensure that unique dummy blocks are available for each access. As in case of the DR-LogSet, the reshuffled versions of the temporary stashes are written to a temporary {\em write-only workspace}. The reshuffled temporary stashes in the workspace replace the old versions in the StashSet after $c$ queries (as part of queries, Section \ref{desc:query}). 

\begin{small}
	\begin{algorithm}
		\footnotesize
		\caption{$\mathsf{readStashSet(id, i)}$}\label{readstashset}
		\begin{algorithmic}[1]
			\State Read $Stash$
			\For{$TempStash_{j} \in StashSet$}
			\State $index\_blk \gets$ Read index for $TempStash_{j}$
			\If {$id \in index\_blk$ and $id$ not already found}
			\State Read $b_{id}$ from $TempStash_{j}$
			\Else
			\State Read $d_i$ from $TempStash_{j}$
			\EndIf 
			\EndFor
			\State $TempStash_{i} \gets$ Read $ith$ temporary stash 
			\State $TempStash_{i}.reshuffle(rand)$
			\State  Write-back $TempStash_{i}$ to {\em temporary workspace}
		\end{algorithmic}
	\end{algorithm}
\end{small}

\paragraph{Querying the StashSet}
$\mathsf{readStashSet}$ (Algorithm \ref{readstashset}) takes as input: i) the logical address $id$ of the block to query, and ii) the client query identifier $i$ and performs the following

\begin{enumerate}
	\item Read current Stash.
	\item {\em Privately query temporary stashes:} Read {\em one} block each from the temporary stashes in {\em descending order} of temporary stash identifiers 
	\begin{itemize}
		\item Determine if $id \in$ index block of $TempStash_{j}$
		\item If $b_{id} \in TempStash_{j}$, then read $b_{id}$
		\item Otherwise, read dummy $d_{i}$.
	\end{itemize}  
	\item Reshuffle temporary stash, $TempStash_{i}$.
\end{enumerate}

\paragraph{Obliviousness}
 $\mathsf{readStashSet}$ ensures that the server does not learn:
i) the identity of the block being queried, and ii) the last access time of the block.

\begin{itemize}
	\item  One block is read each from each of the temporary stashes in a specific order, regardless of the temporary stash that actually contains the required block. This prevents the server from learning when the block was added to the StashSet. Due to the random permutation of blocks, the (encrypted) block read from a temporary stash appears random to the server.

	\item Within a single round of $c$ queries, there can be 
	only one query for a particular block $b$ 
	since if two parallel clients want to access the same block, only one
	client issues a real query while the other client issues a dummy query.

	\item Each stash contains at least $c$ dummy blocks, 
	therefore unique 
	dummy blocks can be read for each of the $c$ accesses in a single query round from a temporary stash before a random reshuffle breaks correlations. 
\end{itemize}

\begin{customthm}{3}
	\label{stashset_lemma}
	The accesses to the StashSet produce transcripts that are indistinguishable
	and independent of the block that is being queried.
\end{customthm}

\begin{proof}
	We show the existence of a simulator that produces indistinguishable access transcripts from $readStashSet(id,i)$ with only public knowledge of i) the corresponding query identifier, which depends on -- i) the number of queries that have started execution prior to this query in the current round, and ii) the blocks that have already been accessed from StashSet  in the current round

	\begin{itemize}
		\item { $\Access^{Real}(readStashSet(id, i))$: The real access transcript produced by $readStashSet(id, i)$}:

		\begin{enumerate}
			\item Read current $Stash$
			\item Read search indices for all stashes in the read-only set
			\item Read {\em one} block from each temporary stash 
			\item Reshuffle the $ith$ temporary stash from the StashSet, when counted in the ascending order of the stash IDs. Add the reshuffled temporary stash to the {\em temporary workspace}.
		\end{enumerate}

		\item { $\Access^{Sim}(readStashSet(id, i))$: Transcript produced by a simulator with only public information}:

		\begin{enumerate}
			\item Read current $Stash$
			\item Read search indices for all temporary stashes in the read-only set
			\item For each stash $TempStash_i$, if $j$ accesses have been performed since $TempStash_i$ was last reshuffled, 
			read a random block from $TempStash_i$ that has not been read within the last $j$ accesses.
			\item Simulate reshuffling the $i^{th}$ temporary stash from the StashSet  
			by creating and adding a new stash to the {\em temporary workspace} containing only dummy blocks.
		\end{enumerate}
		
	\end{itemize}
	
	Due to semantic security of the encryption scheme, 
	\begin{small}
		$$\Access^{Real}(readStashSet(id,i)) \approx_c { \Access^{Sim}(readStashSet(id,i))}$$
	\end{small}
	
\end{proof}

\subsection{Asynchronous Eviction Commits}
\label{desc:async_commit}

\begin{small}
	\begin{algorithm}
		\footnotesize
		\caption{$\mathsf{Evict.AsyncCommit(i)}$}\label{evict_commit_algo_async}
		\begin{algorithmic}[1]
			\State $QueryLock.lock$
			\State Update position map and metadata on data tree path
			\For{$bkt \in EST$}
			\If{$bk.timeStamp <i$}
			\State Copy  $bkt$ from client-side cache to data tree
			\EndIf
			\EndFor
			
			\State Copy remaining buckets on eviction path from write-only tree to data tree
		\State Add $TempStash_{i}$ to StashSet
	\State Clear query log 
	\State Clear bigentry log from DR-LogSet
		\item[] // Replace main stash with the temporary stash if this eviction executed the critical section earliest
			\If{$i < j, \forall j \in StashSet$}
			\State $Stash = TempStash_{i}$
			\EndIf
				\State $QueryLock.Unlock$
		\end{algorithmic}
	\end{algorithm}
\end{small}

 The asynchronous commit protocol (Algorithm \ref{evict_commit_algo_async}) uses the {\em temporary eviction identifier}, $i$, and  performs the following

\begin{enumerate}
	\item Update the position map and metadata on data tree path.
	\item Sync data tree:
	\begin{itemize}
		\item {\em Copy contents of EST buckets from client-side cache to data tree} -- One critical difference here from the synchronous protocol (Algorithm \ref{evict_sync_commit_algo}) is that instead of copying all the EST buckets from the client-side cache to the data tree, only a part of the path may need to be updated based on the state 		of commits.  This is because a subset of the EST buckets may have been already updated by more recent evictions that accessed the critical section
		later, but committed earlier.  These buckets are in their most recent state
		and need not be updated during the commit. Buckets can be identified by storing the  {\em temporary eviction identifier} of the eviction that last updated the bucket as part of the bucket metadata.
		
		\item Copy remaining eviction path from write-only tree to data tree.
	\end{itemize} 
	\item Add temporary stash to the StashSet.
		\item Clear the query log and the bigentry log corresponding to the eviction from the DR-LogSet.
	\item {\em If required, set the temporary stash as the main stash:} If this eviction executed the critical section before all other ongoing evictions, then set the temporary stash as the main stash. This ensures {\em synchronous} updates to the main stash -- evictions update the main stash in the order in which they execute the critical section.
\end{enumerate}

\paragraph{Commit Cost} 
	\begin{itemize}
		\item {\em Updating metadata:} Due to the small metadata size ($\O(\log ^{2}N)$ {\em bits} \cite{ringoram}), updating metadata along the eviction path in the data tree has an overall asymptotic access complexity of $\O(\log N)$ blocks.
		
		\item {\em Upadating position map:} With the position map stored in a PD-ORAM \cite{privatefs}, an update has an overall asymptotic access complexity of $\O(\log{N})$ blocks.
		
		\item {\em Committing changes to EST:} 	Committing changes to the eviction subtree from 	the client-side cache has an overall access complexity $\O((\log{k}+1)\cdot\log N)$ blocks.
		A simple optimization here is to store the contents of the client-side cache on a designated 
	{\em server-side cache}. In that case, committing these changes will only include server-side copies, independent of the network bandwidth.
	
		\item {\em Server-side operations:} The commit also copies
	data and swaps references for several server-side data structures  -- this adds negligible bandwidth overhead.
	
	\end{itemize}


\subsection{Query Protocol}
\label{desc:query}

	\begin{algorithm}[t]
		\caption {$\mathsf{query(id)}$}\label{query}
		\footnotesize
		\begin{algorithmic}[1]
			\State $QueryLock.lock$
				\State Read query log
				\If{$id \in$ query log}
				\State Append $dummy$ entry to query log. // Dummy accesses in next steps
				\Else
				\State Append $id$ to query log.
				\EndIf
			\State $QueryLock.Unlock$	
			\State $i \gets QueryLog.length\%c$ // query identifier
			\State $stash \gets readStashSet(id,i)$
			\State $leafID \gets PM.read(id)$
			\State $path \gets DataTree.readBlocksFromPath(leafID)$
			\item[] // Request order based synchronization
			\State $BlocksFromLogSet \gets readLogSet(id)$
			\State $blk \in BlocksFromLogSet \cup path \cup stash $ // Required block
			\State Update $blk$ for writes
			\State $writeLogSet(blk,i)$
			\If{$i = c-1$}
			\State // Replace all bigentry logs in the DR-LogSet with their reshuffled versions in the temporary workspace
			\State // Replace all temporary stashes in the StashSet with their reshuffled versions in the temporary workspace
			\State Initialize new query log for future queries.
			\EndIf
			
		\end{algorithmic}
			
	\end{algorithm}

 The parallel query protocol (Algorithm \ref{query}) includes:
 \begin{enumerate}
 	\item Read and append entry to the query log.
 	\item Query position map.
 	\item Query StashSet (Algorithm~\ref{readstashset}).
  	\item Read path from data tree.
 	\item {\em Request order based synchronization:} The client waits for previous clients (that registered their query before in the current query round) to finish their queries. 
 	\item Query and update DR-LogSet. (Algorithm \ref{readlogset}, \ref{writelogset})
 \end{enumerate}

\begin{customthm}{4}[Correctness]
	\label{lemma:stashset}
	
	The most up-to-date version of a block is found by queries either on the path indicated 
	by the position map, in the StashSet or in the DR-LogSet.
	
\end{customthm}

\begin{proof}
	First, note that the DR-LogSet does not contain duplicate blocks since when a block is read from a particular log,
	it is effectively removed from there by removing the entry from the search index. During evictions, 
	blocks that do not have corresponding index entries are discarded. Moreover, before an eviction commits, 
	the search index of the corresponding log is checked and position map entries are added
	for only those blocks that haven't been accessed in successive query rounds that were 
	performed while the eviction was in progress.
	Therefore, parallel evictions from the logs 
	in the DR-LogSet update position map entries for unique blocks and  
	the order in which the position map entries are updated is not relevant for consistency.
	
	W.l.o.g. let the StashSet 
	already contain $TempStash_{i-1}$. By induction 
	it can be shown that a similar argument will hold for any $TempStash_{j}, 1< j < i$ 
	in the StashSet.
	
	Consider the scenario 
	where the eviction with identifier $i+1$ commits before eviction $i$.
	Consider a block $x \in TempStash_{i}$. As per the eviction protocol,
	$TempStash_{i}$ contains blocks from $TempStash_{i-i}$, the eviction path 
	$p$ from the data tree  
	and the corresponding bigentry log, $l$ in the DR-LogSet. Thus, 
	the following three cases are possible: $x \in TempStash_{i-1}$,
	$x \in p$ or $x \in l$. As $l$ and $p$ are unmodified as $i$ has not committed and $TempStash_{i-1} \in StashSet$, 
	a client querying for 
	$x$ will find the block in either of the three locations. Also, note that if $x \in l$, then $l$ 
	contains the most updated copy of the block. Thus, $TempStash_{i+1}$ is added to the StashSet without
	overwriting the main stash .
	
	If on the other hand, $i$ commits before $i+1$, this follows the order in which the evictions execute the critical section. This reduces to the synchronous commit case which trivially ensures the theorem.
	
	%
	
	%
\end{proof}

 \begin{customthm}{5}[Query Privacy]
 	The query protocol (Algorithm \ref{query}) produces
 	indistinguishable access transcripts that are independent of the item being accessed (Definition 1).
 \end{customthm}
 
\begin{proof}
	In order to demonstrate that the query access 
	transcripts are independent of the underlying queries, 
	we show that the  same transcript can be 
	generated by a simulator using only 
	public information without any knowledge of the underlying query. This uses results from 
	Theorem \ref{drl_lemma} and Theorem \ref{stashset_lemma}.

	\begin{itemize}
		\item $\Access^{Real}(query(id))$: The read access transcript produced by $query(id)$: 
		
		\begin{enumerate}
			\item Append $id$ to QUERY LOG
			\item Read QUERY LOG
			\item $i \gets query_identifier$
			\item $\Access^{Real}(readStashSet(id, i))$
			\item Query Position Map (PD-ORAM)
			\item Read dummy and real blocks from path $p$ in data tree
			\item REQUEST ORDER BASED SYNCHRONIZATION -- client must wait till all previous clients finish Step 8
			\item  $\Access^{Real}(readLogSet(id))$
			\item $\Access^{Real}(writeLogSet(blk,id))$
		\end{enumerate}
		%
		%
		%
		
		\item $\Access^{Sim}(query(id))$: Transcript produced by a simulator with only knowledge  of the blocks accessed since a log $l_i$ has been previously reshuffled:
		
		\begin{enumerate}
			\item Append a 
			random encrypted entry to the query log 
			\item Read the query log
			\item $\Access^{Sim}(readStashSet(id, i))$ 
			\item Query position map for a dummy item
			\item Read only dummy blocks from a random path in data tree
			\item REQUEST ORDER BASED SYNCHRONIZATION -- client must wait till all previous clients finish Step 8
			\item  $\Access^{Sim}(readLogSet(id))$ 
			\item $\Access^{Sim}(writeLogSet(blk,id))$ 
		\end{enumerate}
		
	\end{itemize}
	
	Due to Theorem \ref{drl_lemma}, Theorem \ref{stashset_lemma} and semantic security of the encryption scheme, 
	
	$$\Access^{Real}(query(id))) \approx_c { \Access^{Sim}(query(id))}$$
	
\end{proof}

\paragraph{Resolving Conflict Between Queries and Commits}
Note that both the asynchronous commit protocol (Algorithm \ref{desc:async_commit}) and the query protocol (Algorithm \ref{desc:query}) must first acquire the {\em query lock} in Step 1. Without explicit synchronization between the clients, this can lead to race conditions and indefinite waits. To resolve this, \sysname~ enforces two simple policies 

\begin{itemize}
	\item {\em Commits do not preempt queries.~} Once a query starts execution by registering an entry in the query log in Step 2, it completes all the steps before an eviction commit can acquire the {\em query lock}. Specifically, when an eviction wants to commit, it checks the number of queries that have started execution (size of the query log) and the number of queries that have completed (size of the current data result log). If there are pending queries, the eviction waits for them to finish before acquiring the query lock.
	\item {\em Queries wait for commits to bound the DR-LogSet.~} As a result of the above condition, eviction commits may have to wait indefinitely. To prevent this, before a query begins execution it checks the number of bigentry logs pending eviction in the DR-LogSet. If the the DR-LogSet is full (already contains $c$ bigentry logs), it waits for a commit.
\end{itemize}

\paragraph{Query cost}
In addition to a position map query ($\O(\log N)$ blocks) and reading a block from each bucket along a path in the data tree ($\O(\log N)$ blocks), 
queries also read

\begin{enumerate}
	\item The current stash of size $\mathsf{MaxStashSize}$ blocks.
	\item The current DRL of size $c$ blocks.
	\item {\em One} block from each DR-LogSet bigentry log.
	\item {\em One} block from each StashSet temporary stash.
	\item {\em One} bigentry log to reshuffle.
	\item {\em One} temporary stash to reshuffle. 
\end{enumerate}

$\mathsf{MaxStashSize} \in \O(\log N)$ blocks \cite{ringoram} and as both the DR-LogSet and StashSet can contain a maximum of $c$ bigentry logs/temporary stashes, $2\cdot c$ blocks are read in total for steps 3 and 4. Each bigentry log contains $2\cdot c$ blocks and each temporary stash has $\mathsf{MaxStashSize}+c$ blocks. Thus, Step 5 and 6 have an overall access complexity of $\O(\log N + c)$
blocks. The overall 
query access complexity is $\O(\log N + c)$ blocks.


\section{Experiments}
\label{oram:evaluation}
 
\paragraph{Implementation}
 \sysname~ has been implemented in Java. We thank the authors of TaoStore \cite{taostore}
and PrivateFS \cite{privatefs} for providing their implementations for comparison.

\paragraph{Experimental Setup}
For all experiments, the server runs on a storage optimized i3.4xlarge Amazon EC2 instance (16 vCPUs and 4x800 GB SSD storage). Clients/proxy run on:
\begin{itemize}
	\item Bandwidth-constrained scenario: 4 Linux machines on the local network, with Intel Core i7-3520M CPU running at 2.90GHz, 16GB DRAM. The server-client bandwidth  is measured as 7MB/s using iperf \cite{iperf}.
	\item High-bandwidth scenario: Clients run on 
	run on t2.xlarge instance with 4 vCPUs and 
	16GB of RAM, within the same VPC as the server. In this case, the link bandwidth is measured to be 115MB/s. 
	  
\end{itemize}

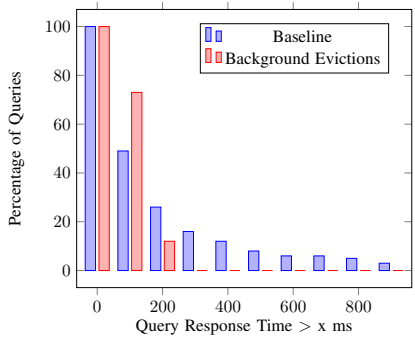
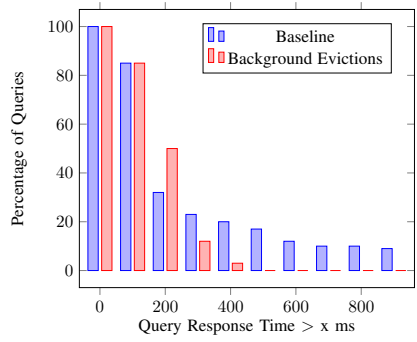
\begin{figure*}[ht!]
	\centering
	\subfigure[Query response time distribution. Evictions after 8 queries ($c = 8$) ]{\label{qrd_8}
		\begin{tikzpicture}[scale=0.65]
		\begin{axis}[
		x tick label style={
			/pgf/number format/1000 sep=},
		ylabel=Percentage of Queries,
		xlabel=Query Response Time $>$ x ms,
		enlargelimits=0.07,
		legend style={at={(0.65,0.95)},
			anchor=north,legend columns=1},
		ybar,
		bar width=6pt
		]
		\addplot 
		coordinates {(0,100) (100,49)
			(200,26) (300,16) (400,12) (500,8) (600,6) (700,6) (800,5) (900,3)};
		
		\addplot 
		coordinates {(0,100) (100,73)
			(200,12) (300,0) (400,0) (500,0) (600,0) (700,0) (800,0) (900,0)};
		
		\legend{Baseline,Background Evictions}
		\end{axis}
		\end{tikzpicture}
		}
	\hspace{0.5cm}
	\subfigure[Query response time distribution. Evictions after 32 queries ($c = 32$) ]{\label{qrd_32} 
		\begin{tikzpicture}[scale=0.65]
		\begin{axis}[
		x tick label style={
			/pgf/number format/1000 sep=},
		ylabel=Percentage of Queries,
		xlabel=Query Response Time $>$ x ms,
		enlargelimits=0.07,
		legend style={at={(0.65,0.95)},
			anchor=north,legend columns=1},
		ybar,
		bar width=6pt
		]
		\addplot 
		coordinates {(0,100) (100,85)
			(200,32) (300,23) (400,20) (500,17) (600,12) (700,10) (800,10) (900,9)};
		
		\addplot 
		coordinates {(0,100) (100,85)
			(200,50) (300,12) (400,3) (500,0) (600,0) (700,0) (800,0) (900,0)};
		
		\legend{Baseline,Background Evictions}
		\end{axis}
		\end{tikzpicture}
			}
	\vspace{-0.3cm}
	\caption{\footnotesize (a) The baseline implementation with serial queries and evictions has higher query response times as queries are blocked during evictions. Background evictions bound query response time better. (b) Background evictions remain largely unaffected by the frequency of eviction and help to upper-bound query response times better as evictions become more expensive to support higher eviction frequency}
\end{figure*}

\begin{figure*}\hspace{0.2cm}
		\subfigure[Query Througput (ops/s). Higher is better. B/W = 7MB/s. DB size = 20GB. ] {
			\label{throughput}
		\begin{tikzpicture}[scale=0.65]
		\begin{axis}[
			legend style={at={(1.05,-0.25)},
				anchor=east,legend columns=-1},
				xlabel=Number of Parallel Clients,
				ylabel=Query Throughput
		]
		\addplot coordinates {(1,4.8) (5,19.2) (10,35.7) (15,59.7) (20,65.8) (25,68) (30,67.6)};
		
		\addplot coordinates {(1,13.73) (5,28.42) (10,41.14) (15,40.4) (20,39.9) (25,38.837) (30,38.65)};
			
		\addplot coordinates {(1,1.3) (5,2.7) (10,4.8) (15,5.6) (20,5.9) (25,5.8) (30,5.9)};
				\legend{\sysname~,TaoStore,PD-ORAM}
		
		\end{axis}
		\end{tikzpicture}
		}
	\subfigure[Query Response Time (in ms). Lower is better. B/W = 7MB/s. DB size = 20GB. 
]{ \label{qr}
	\begin{tikzpicture}[scale=0.65]
	\begin{axis}[
			legend style={at={(1.05,-0.25)},
				anchor=east,legend columns=-1},
		xlabel=Number of Parallel Clients,
		ylabel=Query Response Time
			]
	
	\addplot coordinates {(1,712) (5,714) (10,720) (15,725) (20,731) (25,750) (30,770)};

	\addplot coordinates {(1,375) (5,451) (10,517) (15,626) (20,740) (25,825) (30,875)};

	\addplot coordinates {(1,1500) (5,1502) (10,1500) (15,1500) (20,1507) (25,1510) (30,1507)};
		\legend{\sysname~,TaoStore,PD-ORAM}
	
	\end{axis}
	\end{tikzpicture}
}
	\subfigure[Query Throughput (ops/s). Higher is better. B/W = 115MB/s. DB size = 20GB. ]{  
		\label{throughput_high}
		\begin{tikzpicture}[scale=0.65]
		\begin{axis}[
		legend style={at={(0.90,-0.25)},
			anchor=east,legend columns=-1},
		xlabel=Number of Parallel Clients,
		ylabel=Query Throughput
		]
		
		\addplot coordinates {(1,8.1) (8,71.2) (16,135) (24,198.8) (32,220) (40,250) (48,277) (56,298) (64,300)};

		\addplot coordinates {(1,17.2) (8,91) (16,162) (24,200) (32,225) (40,224) (48,223) (56,224) (64,225)};

		\legend{\sysname~,TaoStore}
		
		\end{axis}
		\end{tikzpicture}
	
		}
	
\vspace{-0.3cm}
\caption{\footnotesize (a) TaoStore query throughput plateaus at 10 clients due to proxy bandwidth limitations. ConcurORAM overall throughput scales gracefully up to 30 clients achieving a max. query throughput of 65 ops/s.
(b) Query response time for TaoStore increases almost linearly with increasing clients as queries contend for fixed number of proxy threads. Queries from different clients in \sysname~ and PD-ORAM are independent and thus the query response time remains unaffected by increasing number of clients. (c) With higher bandwidth \sysname~ can achieve higher overall throughput and plateau only when reaching the server-side limits.}
\vspace{-0.5cm}

\end{figure*}
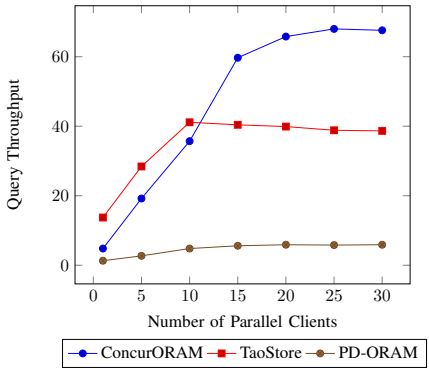
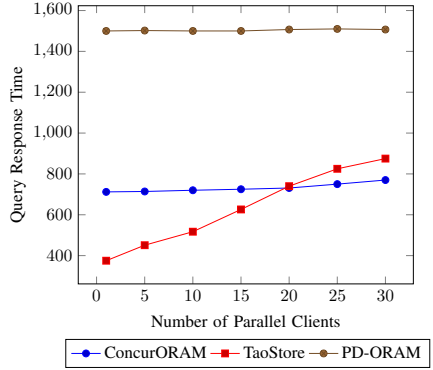
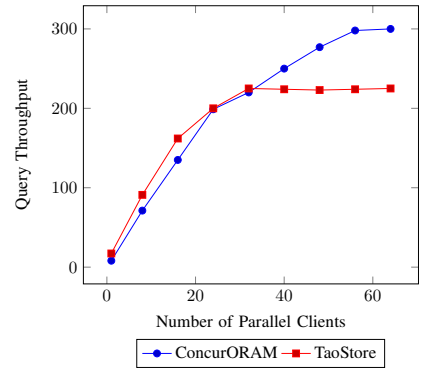

\subsection{Asynchronous Accesses for Stateless Client}
To better understand the benefits of asynchronous accesses in a stateless client setting, we first implemented a version of \sysname~ with serial queries but equipped with the parallel background eviction support, and compared against a baseline implementation with no background evictions. All metadata (position map, stash etc.) are outsourced to the server.

The parameter of interest is the distribution of query response times. With serial queries, periodic evictions block queries for significant periods of time. In fact, less frequent evictions are not helpful, since this results in a proportional increase in bandwidth due to enlarged buckets.

Figure~\ref{qrd_8} and \ref{qrd_32} compare the distribution of query response times for \sysname~ with parallel evictions, against the baseline where evictions are performed serially, blocking the queries. 
Background evictions bound the query response times better while performance is severely affected by intermittent blocking evictions for the baseline.


\subsection{Parallel Queries with Multiple Clients}
We compare parallel query throughput and latency with prior work \cite{privatefs,taostore}. Although TaoStore \cite{taostore} operates in a different trust model, we compare nonetheless 
to demonstrate the limitations of having a centralized proxy. 

If the proxy is over-provisioned, or close to the server (such as within the same Amazon VPC), TaoStore is capable of achieving a high query throughput. But this is seldom the case -- an enterprise deploying a (trusted) proxy to route queries will be more likely to place the proxy within its own protected network, rather than deploying it in the same network as the untrusted server, which will at least require trusted execution guarantees etc. Further, placing the proxy near the server will improve proxy performance, but will not necessarily translate to better query throughput for the clients, which will now be constrained by their own link bandwidths with the proxy.

\paragraph{Query Throughput}
Our experiments clearly demonstrate this phenomenon. Figure~\ref{throughput} shows the overall throughput and the Figure \ref{qr} shows the average query access times for up to 30 parallel clients for \sysname~, TaoStore and PD-ORAM. Clients/proxy are run within our local network, and are thus subject to realistic bandwidth constraints while interfacing with the EC2 server. To conduct experiments with a reasonable number of physical machines, we deploy up to 8 client threads from each of the aforementioned local machines.

Both TaoStore and \sysname~ outperform PD-ORAM due to an asymptotically more efficient (by a factor of $\O(\log N)$) base ORAM. Due to the bandwidth limitations of the TaoStore proxy, the throughput for TaoStore plateaus at 10 clients. Expectedly, \sysname~ can support upto 30 clients without throttling the throughput. At this stage, each of the client machines runs an approximate 8 threads, utilizing the full link bandwidth. Clearly, with independent machines, an even greater number of clients can be supported.

\paragraph{Query Response Time}
The query response time for both PD-ORAM and \sysname~ remain almost constant as clients can query independently. On the other hand, the query response time for TaoStore increases almost linearly with increasing clients as multiple client queries need to contend for the fixed number of query threads deployed by the proxy. At 20 clients, the query response time for TaoStore surpasses the query response time for \sysname~. Note that the higher initial query response time for \sysname~ is because of the stateless design -- clients must fetch all metadata including the position map entries and the stash from the server, while TaoStore benefits from storing all metadata on the proxy.

\paragraph{High Bandwidth Scenario}
With higher available bandwidth, e.g., when clients and server are within the same network, both \sysname~ and TaoStore benefit from appreciable overall throughput increase. (Figure \ref{throughput_high}). Even in this setting, \sysname~ supports a larger number of clients with a higher overall throughput compared to TaoStore. In fact, as we measure, the plateau observed at around 60 clients, is primarily because the large number of client threads throttle the server-side compute (CPU utilization at 100\%) instead of the bandwidth. With a more compute-optimized server, \sysname~ can support a higher number of parallel clients. 

\paragraph{Choice of Parameters}
In the experiments above, we set the query round size $c=$ number of parallel clients. Further, we set the number of consecutive eviction that can execute in parallel, $k = c$. This ensures that all bigentry logs in the DR-LogSet can be evicted to the tree in parallel. Since the value of $c$ impacts query access complexity (Section \ref{desc:query}), the overall through plateaus when increasing $c$ (as shown by Figure 6).
 
In general, $c$ should be determined experimentally based on
network conditions to achieve maximum throughput. With a
large number of parallel clients, it is possible that maximum
throughput is achieved when $c <$ number of parallel clients.
In this case, some parallel queries may need to wait for
the completion of one (or more) query rounds. 

\section{Extensions}
\label{analysis}

\subsection{Fault Tolerance}
In the following, we detail how \sysname~ can proceed gracefully in the event of system crashes, network failures etc.

\paragraph{Crashes During Queries}

\begin{enumerate}
	\item {\em Updating the query log (Algorithm \ref{query}, Steps 1 - 4):} If the client fails while updating the query log, it will do so while holding the query lock. After a specified timeout, other clients can simply proceed with their queries. 
	\item {\em Prior to adding queried block to DRL (Algorithm \ref{query}, Steps 5 - 12):} In this case, the client has not yet updated the current data result log with the result of its query. Recall that clients that started execution later wait for this result. After a specified timeout, the {\em next} client waiting for access to the data result log can repeat the query on behalf of the failed client.
	
	Observe that this does not leak privacy -- if a query fails, the {\em next} client (public information),  always repeats the same query after a specified time, resulting in exactly the same access pattern. 
	
	\item {\em Reshuffling the bigentry log and temporary stashes:} In this case, the failed client will not write back to the temporary workspace. The client that replaces the old versions with the reshuffled version at the end of a query round, can perform the required reshuffles.
\end{enumerate}

\paragraph{Crashes During Evictions}
 Evictions can fail at various stages, and although specified timeouts will allow the protocol to proceed, it is important to know whether to {\em roll back} changes and start afresh, or to {\em roll through} and continue with the protocol for the sake of consistency. 

\begin{itemize} 
	\item {\em Stage 1:} If the eviction fails in stage 1, which only includes read-only access to the write-only tree, a different client can takeover and start afresh.
	
	\item {\em Stage 2 (critical section):} In stage 2, the eviction will fail while holding the {\em processing lock}. As a result, subsequent evictions waiting for results from the critical section will not proceed. In case of a crash, the eviction can be restarted by another client after rolling back the changes performed in the critical section. Effectively this means disregarding any changes made in the critical section (to the EST buckets and temporary stash) by the failed client and starting afresh.
	
	\item {\em Stage 3:} At this stage, results generated in the critical section by the failed evictions may have been used by subsequent evictions entering the critical section. Thus, the changes cannot be rolled back. Instead, a different client can continue with the eviction process. This requires one critical piece of information -- {\em the pseudo-random mapping} used by the failed eviction for mapping blocks in the bigentry logs to new paths in the data tree. While in the critical section, this information is added to the {\em eviction log}, and can be used later by a different client for completing the eviction. 
	
	\item {\em Commits:} Another client can perform the commit on behalf of the failed client with some additional information 
	
	\begin{itemize}
		\item {\em Server-side cache:} Firstly, all contents of the client-side cache which include the temporary stash and the eviction sub-tree buckets needs to be stored on the server. This can be used by a different client for committing updates, in case of a crash.
		
		\item {\em Eviction metadata:} The client that performs the eviction on behalf of the failed client has to perform several other tasks. This includes updating the position map for blocks evicted, updating metadata on data tree path, and clearing query and data result logs.

		To perform these tasks, the client requires the following additional {\em metadata}, which can be stored in the {\em eviction log} while in the critical section: i) the eviction identifier, ii) the eviction path identifier, iii) 
		the logical identifiers of the evicted blocks {\em and} the paths to which they are mapped in the data tree, and iv) the identifiers for the query log and the data result log.  
		
	\end{itemize}
\end{itemize}

\subsection{Security Against Malicious Server}
%
As a first step, integrity of the server-side data structures can be ensured using existing techniques e.g., {\em embedding a Merkle tree in the ORAM tree} \cite{pathoram}, storing HMACs over the DR-LogSet, StashSet etc. In a single client settings, the root of the Merkle tree and the HMACs can be cached client side and verified for each access. Unfortunately, in a multi-client setting, these variables need to be stored on the server for consistency, leaving open the possibility of replay attacks.

Specifically, as pointed out by previous work \cite{privatefs}, the server could simply replay the query log, presenting different views of ongoing transactions to concurrent clients. As a result, clients querying for the same block will have the same resulting access pattern to the server data structures, leaking inter-client privacy. The server could also replay contents of the server-side data structure along with the Merkle tree root hashes, HMACs etc.
Without inter-client communication, the best we can do to prevent this is {\em fork consistency} \cite{sundr} -- if the server selectively replays the states of server-side data structures and presents different (possibly conflicting) views of the system to different clients, then these views cannot be undetectably unified later.

\paragraph{Protecting Against Query Log Replays} Similar to previous work \cite{privatefs}, the key idea here is to store a hash tree over all previous queries on the server. Specifically, for each query, a client updates the hash tree with a new leaf record, which includes: i) the clients' own unique identifier, and ii) the logical address of the block queried.

This record (and the updated root hash) is stored client side and as part of its {\em next} query, the client verifies that the record exists in the hash tree.  Since hashes are unforgeable, the server cannot add a new record to the hash tree and forge the root hash value in case of forking attacks. Further, the clients' unique identifiers being included in the record ensures that even if two clients query for the same block, or update the same data structure contents (as a result of a forking attack), the hash records they generate will be different with very high probability, when using a collision resistant hash function.


\paragraph{Protecting Against Data Structure Replays} Similarly, to protect against replay of data structure states, a server-hosted hash tree can record updates to the {\em data structure integrity variables}. 
This specifically includes the Merkle tree root hash for the data tree, and HMACs for the DR-LogSet and StashSet.

The client performing updates to the corresponding data structure stores the record (and root hash) locally and verifies that the record exists in the hash tree for subsequent accesses. Using the hash tree, a client can verify the integrity of the data structures. If the server replays states, the client which performed the last update will have a different view of the data structures (root hashes, HMAC values) from the other clients in the system. These views cannot be merged later on by the server without being detected.



\section{Conclusion}
\label{oram:conclusion}

\sysname~ is a  multi-client  ORAM that eliminates
waiting for concurrent clients and allows overall throughput to scale gracefully
with an increase in the number of clients, {\em without requiring trusted proxies or direct inter-client coordination}.
 This is based on the insight that only
a subset of the server-hosted data structures require parallel access with
privacy guarantees and everything else can be implemented safely as
simple and efficient oblivious data structures.
 \sysname~ benefits from a novel eviction protocol that enables multiple concurrent clients to evict asynchronously, in parallel (without compromising
consistency), and in the background without blocking
ongoing queries.



\scriptsize
\bibliographystyle{abbrv}
\bibliography{bib/references.bib}
\end{document}